\documentclass{aa}  
\usepackage{graphicx}
\usepackage[T1]{fontenc}
\usepackage[utf8]{inputenc}
\usepackage{CJKutf8}
\usepackage[absolute,overlay]{textpos}
\usepackage{subfig}
\usepackage{natbib}
\usepackage{caption}
\usepackage{rotating}
\usepackage{threeparttable}
\usepackage{titlesec}
\usepackage{placeins}
\usepackage{lscape}
\usepackage{longtable}
\usepackage{needspace}
\usepackage{afterpage}
\usepackage{tabularx}
\usepackage{ltxtable}
\usepackage{float}
\usepackage{txfonts}

\usepackage{xcolor}

\usepackage{hyperref}
\hypersetup{
    colorlinks=true,    
    linkcolor=red,     
    citecolor=blue,     
    filecolor=magenta,  
    urlcolor=red       
}
\usepackage{orcidlink}

\begin{document}
\begin{CJK}{UTF8}{gbsn}

\title{Environmental effects as a key factor in shaping star-forming S0 galaxies}

\author{
Pei-Bin Chen \begin{CJK*}{UTF8}{gbsn}(陈培彬)\end{CJK*}\orcidlink{0009-0002-4905-263X},$^{1}$
Jun-Feng Wang \begin{CJK*}{UTF8}{gbsn}(王俊峰)\end{CJK*}\orcidlink{0000-0003-4874-0369},$^{1}$\thanks{E-mail: jfwang@xmu.edu.cn}
Yan-Mei Chen \begin{CJK*}{UTF8}{gbsn}(陈燕梅)\end{CJK*},$^{2}$
Xiao-Yu Xu \begin{CJK*}{UTF8}{gbsn}(许啸宇)\end{CJK*}\orcidlink{0000-0003-0970-535X},$^{3,4}$ Tian-Wen Cao \begin{CJK*}{UTF8}{gbsn}(曹天文)\end{CJK*}\orcidlink{0000-0002-1335-6212}$^{1}$
\\}
\institute{$^{1}$ Department of Astronomy, Xiamen University, Xiamen, Fujian 361005, China\\
$^{2}$ Department of Astronomy, Nanjing University, Nanjing 210093, China\\
$^{3}$ School of Astronomy and Space Science, Nanjing University, Nanjing 210023, China\\
$^{4}$ Key Laboratory of Modern Astronomy and Astrophysics, Nanjing University, Nanjing 210023, China
}

\date{Accepted XXX. Received YYY; in original form ZZZ}

\abstract
{The origins of lenticular galaxies (S0s) can be classified into two main categories: ``minor mergers" in low-density environments (LDEs) and ``faded spirals" in high-density environments (HDEs). The transitional phase in the evolution of S0s, namely, star-forming lenticular galaxies (SFS0s), can serve as an important probe for analyzing the complex processes involved in the transformation between different galaxy types and the quenching of star formation (SF).}
{We attempt to find the impact of different environments on the global properties and spatially resolved quantities of SFS0s.}
{We selected 71 SFS0s from the SDSS-IV MaNGA Survey, comprising 23 SFS0s in HDEs (SFS0s$\_$HE) and 48 SFS0s in LDEs (SFS0s$\_$LE). We examined the effects of the environment, by studying the global properties, concentration index, and radial profiles of the derived quantities.}
{The varied environments of SFS0s do not lead to any significant difference in global properties (e.g., S$\acute{\rm e}$rsic index). By calculating $CI_{\rm H_{\alpha}/cont}$, we observe that different environments may cause varying concentrations of SF. Specifically, SFS0s$\_$LE, affected by external gas mergers or inflow, exhibit a more centrally concentrated SF (i.e., larger $CI_{\rm H_{\alpha}/cont}$). This trend is further supported by $CI_{\rm SFR, H_{\alpha}}$, which only considers the gas disk of the galaxy. This observation is aligned with the observed shrinking of gas disks in galaxies affected by ram-pressure stripping in HDEs. Furthermore, their $\Sigma_{\rm SFR}$ or resolved sSFR are comparable. On average, SFS0s$\_$LE display significantly higher values for both quantities. Finally, the observed D$_{\rm n}4000$ and gas-phase metallicity gradient correspond well to their assumed origins. However, we did not find a significantly lower gas-phase metallicity in SFS0s$\_$LE.}
{We suggest that different environments (i.e., origins) do not have a significant impact on the global properties of SFS0s, but they do indeed affect the distribution of SF. Considering the size of our sample and the unique nature of the galaxy, additional atomic and molecular gas data may provide further details to improve our understanding of these systems.}


\titlerunning{Environment Effects of SFS0s}
\authorrunning{Chen et al.}

\maketitle

\section{Introduction}
\label{introduction}

Substantial efforts have been devoted to understanding the correlation between galaxies and their environments \citep[e.g.,][]{2013pss6.book..207V, 2022MNRAS.515.5877K, 2023MNRAS.521.1292P}. Numerous studies have shown that the environments of galaxies affect their color, gas content \citep[e.g.,][]{2006PASP..118..517B}, and galaxy interactions, which then affect their morphology \citep[e.g.,][]{1980ApJ...236..351D}, current star formation \citep[SF; e.g.,][]{2002MNRAS.334..673L, 2013pss6.book..207V}, and SF history \citep[SFH; e.g.,][]{2012ApJ...746...90A}. Galaxies can be broadly classified based on morphology into elliptical galaxies (EGs), spiral galaxies (SGs), and the intermediate form known as lenticular galaxies \citep[S0s;][]{1936rene.book.....H}. In general, EGs and S0s are referred to as early-type galaxies \citep[ETGs; e.g.,][]{2023RAA....23a5005C}, while SGs are referred to as late-type galaxies (LTGs). Observations show that quenched galaxies tend to have early-type morphology \citep[e.g.,][]{2017MNRAS.471.2687B}. This is not surprising, as ETGs are typically massive galaxies in high-density environments \citep[HDEs;][]{1980ApJ...236..351D}. The concept that ETGs are passive and red death systems is widely accepted \citep[e.g.,][]{2001AJ....122.1861S, 2022MNRAS.509.1237X}. However, recent observations indicate that some ETGs contain atomic gas, even dust and/or molecular gas. Their star formation (SF) activities have been revealed through far-ultraviolet (FUV) and infrared (IR) radiation \citep[][]{2014MNRAS.444.3427D} or emission-line diagnostic analysis \citep[e.g.,][]{2016ApJ...831...63X, 2023RAA....23a5005C}.

Many works are dedicated to studying the origin of blue ETGs. \citet{2009MNRAS.396..818S} found a sample of 204 blue ETGs exhibiting mild to moderate star formation rates (SFRs) between 0.05 $\sim$ 50 $M_{\odot}$yr$^{-1}$. They discussed the possible positions that blue ETGs may occupy in the overall evolutionary picture. Moreover, they found that their gas-phase metallicity is all supersolar, which is consistent with the mass-metallicity relation \citep[][]{2004ApJ...613..898T}. \citet{2009AJ....138..579K} identified a population of morphologically defined EGs/S0s lying on the locus of LTGs in the color-stellar mass (M$_{\ast}$) space (the blue sequence) at the present epoch. They studied the proportions of these galaxies at three mass scales (i.e., shutdown mass, bimodality mass, and threshold mass) of interest. Moreover, \cite{2010ApJ...725.2312O} used large-scale dark matter simulation to recover the observational result of ``archaeological downsizing," indicating that ETGs have a two-phase formation process: in situ SF and external accretion. Of course, earlier studies have also provided empirical evidence prior to this \citep[e.g.,][]{1998MNRAS.297..427A, 2004cbhg.sympE..36L} and a large amount of observational evidence has been obtained to support this claim \citep[e.g.,][]{2011ApJS..195...15D, 2011MNRAS.413.2943F, 2013MNRAS.432.1862C}. Specifically, \citet{2013MNRAS.432.1862C} used the volume-limited and nearly mass-selected ATLAS$^{\rm 3D}$ sample of 260 ETGs to study their distributions in the mass-size and mass-$\sigma$ planes. They also discovered the three characteristic mass scales and found evidence of this ``two-phase" character (see their Figs. 15 and 16). As one of the components of ETGs, S0s have attracted much attention \citep[e.g.,][]{1972ApJ...176....1G, 2004ApJ...617..867D, 2006ApJ...651..811B, 2015A&A...579L...2Q, 2022MNRAS.515..201C}.
 
Generally, S0s exhibit a central bulge surrounded by a disk but lack prominent spiral arms. Interestingly, \cite{2017MNRAS.471.2687B} found that S0s are quite abundant in various environments (see their Fig. 12) and their specific SFR (sSFR) spans several orders of magnitude (see their Fig. 6), from the red sequence (red sequence galaxies, RSGs) to the green valley (green valley galaxies, GVGs) to the blue cloud (star-forming galaxies, SFGs). Numerous studies have revealed significant differences in the characteristics of S0s \citep[photometric, spectroscopic, and kinematic; e.g.,][]{2013MNRAS.432..430B} in different environments and cover a certain range \citep[][]{2018MNRAS.481.5580F}, suggesting multiple and complex formation pathways. So far, the formation pathways of S0s can be divided into two major categories (faded spiral and minor merger). The variation in the fraction of S0s in different environments is exactly the opposite of the variation in SGs \citep[][]{1980ApJ...236..351D, 1997ApJ...490..577D, 2017MNRAS.471.2687B}. Thus, considering the similar structure and kinematic characteristics \citep[$V_{\rm rot}/\sigma$ $>$ 1;][]{2020MNRAS.492.2955C, 2020MNRAS.498.2372D, 2021MNRAS.508..895D} between the two types, it is generally believed that S0s originate from the fading of SGs. Investigations into S0s across diverse environments have unveiled that those with significant gaseous emission are primarily located in the field and on the outskirts of galaxy clusters, suggesting that the gas either undergoes stripping when galaxies fall into galaxy clusters or the increased gas content results from the infall or merger of external fresh gas \citep[e.g.,][]{2021MNRAS.508..895D}. This hypothesis finds support in various studies \citep[e.g.,][]{1972ApJ...176....1G, 2009MNRAS.394.1991B, 2012ApJS..198....2K, 2018MNRAS.478..351M, 2020MNRAS.492.2955C, 2020MNRAS.498.2372D, 2021MNRAS.508..895D, 2022A&A...659A..46B}, suggesting that it is more prevalent in HDEs such as galaxy groups and clusters, often referred to as ``faded spirals," where SGs lose their gas and SF is rapidly quenched.  

However, this pathway does not explain the apparent difference between the observed features of the S0s and the expected properties of ancestral SGs. Investigations into S0s in low-density environments (LDEs, fields) have unveiled distinct properties \citep[e.g.,][]{2018MNRAS.477.2030D, 2022MNRAS.515..201C, 2020MNRAS.498.2372D}. For instance, S0s in LDEs exhibit redder bulges in comparison to their disks \citep[][]{2017MNRAS.466.2024T}. \citet{2020MNRAS.492.2955C} compared the spatially resolved kinematics of 21 S0s from an extreme environment with the overall stellar populations (SPs) and found that S0s in the field exhibited more pressure-supported ($V_{\rm rot}/\sigma$ $<$ 1), which indicates that mergers or gas accretions \citep[minor merger;][]{2013MNRAS.432.1862C, 2016ARA&A..54..597C} shape their kinematic characteristics. \citet{2020MNRAS.498.2372D} studied 219 S0s from the SAMI Survey and found that minor mergers dominated in isolated galaxies and small groups. Similarly, \citet{2022MNRAS.515..201C} carried out a study on 329 S0s from the SAMI and MaNGA surveys and showed that the minor merger is a viable channel. Furthermore, the presence of decoupled gas kinematics in isolated S0s also supports external sources of gas \citep[][]{2014MNRAS.438.2798K}. In addition, simulations have also supported this channel \citep[e.g.,][]{2015A&A...579L...2Q, 2018A&A...617A.113E, 2021MNRAS.508..895D}. This type of channel is popular in LDEs \citep[e.g.,][]{2021MNRAS.508..895D, 2022MNRAS.515..201C}, such as fields.

Today, studying S0s that are still undergoing SF activity is very interesting. A transitional phase in the evolution of S0s, star-forming lenticular galaxies (SFS0s), serve as important probes for understanding the complex processes involved in the transformation between different galaxy types and the quenching of star formation. \citet{2016ApJ...831...63X} studied 583 S0s from SDSS Data Release 7 (DR7) and found that only 8$\%$ of them showed central SF activity; these active S0s haave lower stellar mass (M$_{\ast}$) and are more inclined to sparse environments. \citet{2022MNRAS.509.1237X} studied 52 star-forming S0s (SFS0s) from the SDSS-IV MaNGA Survey and found the presence of pseudo bulges and different dynamic processes in SFS0s. Moreover, they found that the number of SFS0s in the LDEs is twice that in the HDEs, even though the normal S0s favored the HDEs \citep[][]{1997ApJ...490..577D}. Similarly, \citet{2022MNRAS.513..389R} used sSFR \citep[][]{2014SerAJ.189....1S} as a criterion to study 120 SFS0s from the SDSS-IV MaNGA Survey and found that the SF of these galaxies is center-dominated rather than disk-dominated. These previous works all suggest that SFS0s also have two main formation pathways, as mentioned above. In this work, we accept these two formation pathways in different environments, studying the environmental effects on the global and derivative properties of SFS0s.

We utilized integral field spectroscopy (IFS) data obtained from the Mapping Nearby Galaxies of the Sloan Digital Sky Survey at APO \citep[SDSS-IV MaNGA,][]{2015ApJ...798....7B, 2015AJ....150...19L, 2017AJ....154...28B}. The MaNGA data allow for the analysis of spatially resolved dynamics and chemical compositions in galaxies, providing crucial information about their formation and evolution. Utilizing 17 simultaneous integrated field units (IFUs), each comprising a closely packed array of optical fibers, MaNGA conducted spectroscopic measurements across the surfaces of approximately 10,000 nearby galaxies \citep[][]{2015ApJ...798....7B, 2015AJ....150...19L, 2017AJ....154...28B}. This approach generates 2D maps representing various physical quantities such as stellar and ionized gas velocities, facilitating a comprehensive understanding of the ``life history" of galaxies. These maps provide insights into the birth, assembly, ongoing growth through SF and merger processes, and eventual quenching at later stages \citep[e.g.,][]{2015ApJ...798....7B}. The observed spectrum covers a wavelength range from 3600\,\AA\ to 10300\,\AA, with a spectral resolution of approximately 2000. Notably, the observations span at least 1.5 effective radii (R$_{\rm e}$) for the target galaxy, where R$_{\rm e}$ denotes the radius containing 50$\%$ of the galaxy's light. 

In this work, we constructed a sample of 71 SFS0s in various environments from the SDSS-IV MaNGA survey. Our purposes are: 1) To find out whether SFS0s have different global properties (e.g., S$\acute{\rm e}$rsic index) in different environments; 2) We attempted to establish the relationship between the derivation quantities (e.g., concentration index, D$_{\rm n}$4000, metallicity) of SFS0s and its different formation pathways. This paper is organized as follows: Our sample selection is presented in Sect. \ref{Data}, as well as the environmental information of galaxies. We provide the methodology in Sect. \ref{Methodology}, including extinction correction, the description of morphological parameters, and how to construct radial profiles of derivation quantities. Section \ref{Results} provides our results and the related discussion listed in Sect. \ref{Discussion}. Finally, we list our summary in Sect. \ref{Conclusions}. Throughout this paper, we adopt a set of cosmological parameters as follows: \emph{H$_0$} = 70 km s$^{-1}$ Mpc$^{-1}$ (i.e., \emph{h} = 0.7), $\Omega_{\rm m}$ = 0.30, and $\Omega_{\rm \Lambda}$ = 0.70.


\section{Data and sample selection}
\label{Data}

\subsection{Data}\label{sec2.1}

We utilized the data products from MaNGA \citep[][]{2015ApJ...798....7B, 2016AJ....152..197Y} DR 17 \citep[][]{2022ApJS..259...35A}. The primary data products from MaNGA encompass 3D calibration data cubes generated through the Data Reduction Pipeline (DRP) and 2D maps of derived quantities produced by the Data Analysis Pipeline (DAP) using these cubes. The DAP constructs 2D maps of the derived quantities, including information on stellar, gas, and emission lines, by analyzing individual or binned groups of pixels. The final data cubes provide spectra for each pixel of a target, with each pixel covering an area of 0.$^{\prime \prime}$5 $\times$ 0.$^{\prime \prime}$5. In this work, we utilized the DAP 2D maps (``HYB10-MILESHC-MASTARSSP"\footnote{\url{https://data.sdss.org/sas/dr17/manga/spectro/analysis/v3_1_1/3.1.0/HYB10-MILESHC-MASTARSSP/}}) of the derived quantities. Additionally, MaNGA targets have undergone processing with the Pipe3D IFS data-processing pipeline \citep[][]{2016RMxAA..52...21S, 2016RMxAA..52..171S}. We extracted spatially resolved stellar mass (M$_{\ast}$) and stellar mass surface density ($\Sigma_{\ast}$) from Pipe3D data cubes\footnote{\url{https://data.sdss.org/sas/dr17/manga/spectro/pipe3d/v3\_1\_1/3.1.1/}}. Other intrinsic properties, including redshift, axis ratio (b/a), and R$_{\rm e}$, were sourced from NASA-Sloan Atlas (NSA) catalog\footnote{\url{https://data.sdss.org/datamodel/files/ATLAS_DATA/ATLAS_MAJOR_VERSION/nsa.html}} \citep[v1$\_$0$\_$1,][]{2011AJ....142...31B}. Morphological \citep{2022MNRAS.509.4024D, 2022MNRAS.512.2222V}, environmental \citep{2015A&A...578A.110A, 2015MNRAS.451..660E, 2016ApJ...831..164W} and photometrical \citep{2022MNRAS.509.4024D} information was sourced from the MaNGA value-added catalogs (MaNGA$\_$VAC\footnote{\url{https://www.sdss.org/dr17/data_access/value-added-catalogs/}}).

\subsection{Sample selection}\label{sec2.2}

\begin{figure*}
   \centering
   \includegraphics[width=0.8\textwidth,height=0.55\textwidth, angle=0]{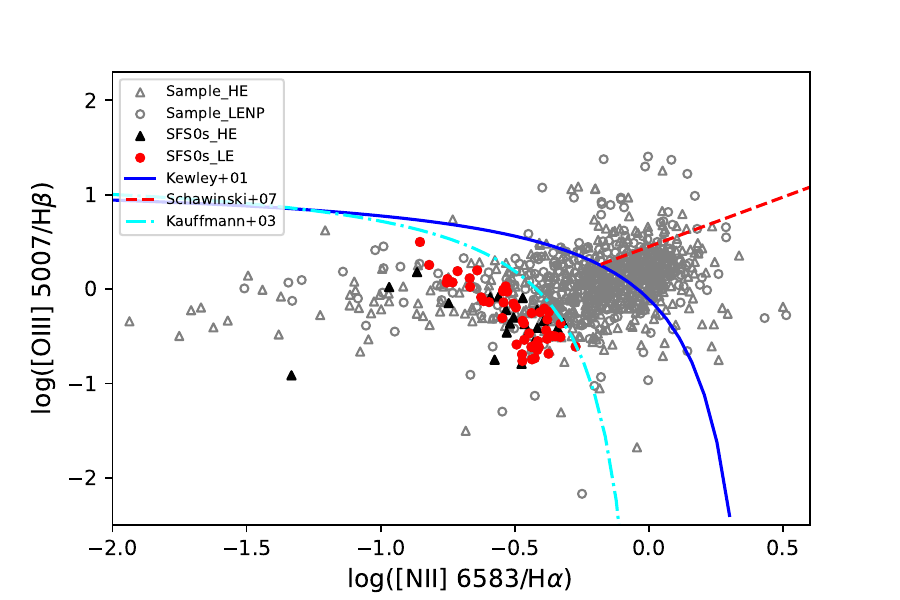}
   \caption{Global BPT diagnosis of our samples. The cyan dotted line, blue solid line, and red dashed line are the theoretical curves from \citet{2003MNRAS.346.1055K}, \citet{2001ApJ...556..121K}, and \citet{2007MNRAS.382.1415S}, respectively. The marks of a triangle in this figure are Sample$\_$HE,  while the marks of a circle represent Sample$\_$LENP. Our SFS0s$\_$HE galaxies are represented by the black-filled triangles, and the red-filled circles represent sample SFS0s$\_$LE. Y-axis is the log([OIII] 5007/H$_{\beta})$, and X-axis is the log([NII] 6583/H$_{\alpha})$, both from Pipe3D v3$\_$1$\_$1 catalog.}
   \label{global-bpt}
   \end{figure*}

MaNGA$\_$VAC provides a Deep Learning Catalogue (hereafter MDLM$\_$VAC) based on the morphological classification of the final MaNGA DR17 galaxy sample \citep[][]{2022MNRAS.509.4024D}. The methods for training and testing the Deep Learning model were described in detail in \citet{2022MNRAS.509.4024D}. These methods are based on a previous work \citep[][]{2018MNRAS.476.3661D}, which provided the classification for about 670,000 objects from SDSS DR7 Main Galaxy Samples. Furthermore, following the method provided by \citet{2022MNRAS.512.2222V}, MaNGA$\_$VAC provides a pure visual morphology classification catalog (hereafter, MaNGA$\_$visual$\_$morpho), covering all galaxies in MaNGA DR17. This classification is derived from the inspection of image mosaics, utilizing a new re-processing of SDSS and Dark Energy Legacy Survey (DESI) images. The classification includes 13 Hubble types and notes the presence of bars and bright tidal debris. Therefore, our selection comprises the following: 

\begin{itemize}
    \item We selected galaxies that pass the basic selection criteria and have ``$T\_Type \leq 0$", ``$P\_LTG < 0.5$", ``$P\_S0 > 0.5$" and ``$VC = 2$" as recommended by MDLM$\_$VAC for identifying S0s. To create a sample containing only S0s, we cross-reference with MaNGA$\_$visual$\_$morpho to ensure that the ``Hubble-type" of the galaxy is S0. Moreover, we excluded the unreliable classification sources (Unsure = 1). Finally, we cross-referenced with the Pipe3D catalog \citep[][]{2020ARA&A..58...99S} and the photometric catalog \citep[][]{2022MNRAS.509.4024D} to obtain a parent sample.
    
    \item Once the parent sample is established, understanding the environmental information of the galaxy becomes crucial. A common method is to measure the projected density to the \emph{N}th nearest group galaxy \citep[e.g.,][]{2018MNRAS.474.2039E}. Information on the local environment of galaxies is provided in the MaNGA$\_$VAC GEMA catalog \citep[hereafter, GEMA$\_$VAC;][]{2015A&A...578A.110A}, where the projected number density parameter ($\eta_{{\rm k},{\rm LSS}}$, i.e., eta$\_$k) of the galaxy is defined as： 

    \begin{equation}\label{eta_k}
        \eta_{{\rm k},{\rm LSS}} = {\rm log}(\frac{\rm k-1}{{\rm Vol}(d_{\rm k})}) = {\rm log}(\frac{3(\rm k-1)}{4\pi d^{3}_{\rm k}})
    \end{equation}
    where $d_{\rm k}$ is the projected physical distance to the kth nearest neighbour, with k is equal to 5 \citep[][]{2004MNRAS.348.1355B, 2016ApJ...831...63X, 2022MNRAS.514.6141J}.
    
    \item By cross-matching the above parent sample with GEMA$\_$VAC, we obtained a sample containing environmental information (hereafter Sample$\_$A). According to the definition of ``eta$\_$k" (formula \ref{eta_k}), Sample$\_$A is divided into two samples: S0s with an eta$\_$k value in HDEs (e.g., galaxy clusters and groups, hereafter Sample$\_$HE) and S0s without an eta$\_$k value in LDEs (e.g., fields, hereafter Sample$\_$LE). Importantly, for Sample$\_$LE, even if they are located in LDEs, some galaxies may make up pairs. Previous works have highlighted the fact that environmental influences on galaxy properties extend beyond cluster cores, affecting all groups with a projection density exceeding 1 galaxy $Mpc^{-2}$ \citep[][]{2002MNRAS.334..673L}. Here, we exclude paired galaxies to obtain a pure LDEs sample (hereafter, Sample$\_$LENP).

    \item We further require that the axis ratio (b/a) is greater than 0.32 (see Sect. \ref{radial profiles}) to avoid the edge-on situation \citep[e.g.,][]{2022MNRAS.509.1237X}. Finally, we employed the emission line flux ($H_{\alpha}$ $\lambda6563$, $H_{\beta}$ $\lambda4861$, [O\,{\sc iii}] $\lambda5007$ and [N\,{\sc ii}] $\lambda6583$) fitted within 1 R$_{\rm e}$ (signal-to-noise ratio, S/N of $\geq$ 3) provided by the Pipe3D catalog \citep{2020ARA&A..58...99S} to perform a global Baldwin, Phillips $\&$ Telervich \citep[BPT,][]{1981PASP...93....5B} diagnosis of our two samples (i.e., Sample$\_$HE and Sample$\_$LENP). This diagnosis helped us to select galaxies located in the H\,{\sc ii} regions (Fig. \ref{global-bpt}). Although active galactic nucleus-host (AGN-host) S0s are excluded in our work, the proportion of AGN-host S0s present in our two parent samples (Sample$\_$HE and Sample$\_$LENP) is close (8$\%$ and 11$\%$), so the bias caused by such screening does not depend on the galaxy's environments, even though different environments have different perturbation mechanisms \citep[][]{2022A&A...659A..46B, 2023A&A...671A.118C}. The slightly higher proportion of AGN-host S0s in LDEs is consistent with previous works \citep[][]{2016ApJ...831...63X}. Furthermore, it is required that these galaxies have an H$_{\alpha}$ equivalent width $>$ 6\,\AA\ within the center of 2.5$^{\prime \prime}$ to ensure that the final targets are SFGs \citep[][]{2020ARA&A..58...99S, 2022MNRAS.509.1237X}.  
\end{itemize}

Following these screening steps, we obtained two parent samples: Sample$\_$HE and Sample$\_$LENP. Then, through the fourth step detailed above, we obtained the final samples for our study: 23 (SFS0s$\_$HE) and 48 (SFS0s$\_$LE). The number of SFS0s in LDEs is about twice that in HDEs, which is consistent with previous results \citep[][]{2016ApJ...831...63X, 2022MNRAS.509.1237X}. The basic information for both samples is listed in Table \ref{tab: composition}, and the detailed information of our targets is listed in Tables \ref{SFS0s_HE} and \ref{SFS0s_LE}. It is worth emphasizing that the final sample (SFS0s) obtained is only a very small part of the S0 population and may not represent all local S0s \citep[e.g.,][]{2022A&A...659A..46B}. In addition, we selected galaxies from a deep-learning catalog based on morphological classification, so there is a certain probability of contamination from SGs. We visually examined the deeper DESI images and found that such contamination was very small (about 8$\%$). Subsequent comparisons have indicated that this is not expected to significantly affect our results (see Sect. \ref{DQ radial profiles}).

\begin{table*}
\caption{Basic information of our samples\label{tab: composition}}
\centering 
\begin{tabular}{cccccc} 
\hline\hline 
Samples & Size & log(M$_{\ast}/M_{\odot})$ $\geq 10.25$ & log(M$_{\ast}/M_{\odot})$ $<$ $10.25$ & S$\acute{\rm e}$rsic index $\geq$ $2$ & S$\acute{\rm e}$rsic index $<$ 2\\
(1) & (2) & (3) & (4) & (5) & (6)\\
\hline 
SFS0s$\_$HE & 23
& 9 (39.1$\%$) & 14 (60.9$\%$) & 10 (43.5$\%$) & 13 (56.5$\%$) \\
SFS0s$\_$LE & 48 & 15 (31.3$\%$) & 33 (68.2$\%$) & 13 (27.1$\%$) & 35 (72.9$\%$) \\
\hline 
\end{tabular}
\\
Notes: In this table, log(M$_{\ast}/M_{\odot})$ = $10.25$ is the division of stellar mass from \citet{2022MNRAS.513..389R} and S$\acute{\rm e}$rsic index of bulges come from the MaNGA-VAC photometry catalog. The percentage in parentheses represents the corresponding proportion of the galaxy.
\end{table*}

\section{Methodology}
\label{Methodology}

\subsection{Extinction correction}

To calculate the derived quantities of each pixel related to SF (e.g., SFR), we take into account dust extinction, a factor considered in many studies \citep[e.g.,][]{2021MNRAS.505..191B, 2022MNRAS.509.1237X}. \citet{2001PASP..113.1449C} estimated the extinction-corrected H$_{\alpha}$ luminosity using the H$_{\alpha}$/H$_{\beta}$ ratio:

\begin{equation}
    f_{\rm i}(\lambda) = f_{\rm o}(\lambda) \times 10^{\rm 0.4E(B-V)_{\rm gas}\kappa^{\rm e}(\lambda)}
\end{equation}
where $f_{\rm i}(\lambda)$ is the intrinsic flux density, $f_{\rm o}(\lambda)$ is the observed flux density, the color excess E(B $-$ V) is:

\begin{equation}
    E(B-V) = 2.15 \times {\rm log[(L_{H_{\alpha}}/L_{H_{\beta}})/(L_{H_{\alpha}}/L_{H_{\beta}})_{\rm i}]}
\end{equation}
Here, L$_{\rm H_{\alpha}}/{\rm L}_{\rm H_{\beta}}$ is the observed ratio, $({\rm L}_{H_{\alpha}}/{\rm L}_{H_{\beta}})_{\rm i}$ = 2.86 is an intrinsic Balmer decrement corresponding to Case B recombination with a temperature of $10^4$ K and electron density n$_{\rm e}$ = $10^2$ cm$^{-3}$ \citep[][]{2001PASP..113.1449C}. Furthermore, $\kappa^{\rm e}(\lambda)$ is the Galactic dust attenuation curve, expressed as \citep[][]{2001PASP..113.1449C}:

\begin{multline}
  \kappa^{\rm e}(\lambda) = 1.17 \times (-1.857 + 1.040/\lambda) + 1.78 \\ {\rm for} \ 0.63\  \rm{\mu m} \leq \lambda \leq 2.20\  \rm{\mu m}
\end{multline}
and:
\begin{multline}
  \kappa^{\rm e}(\lambda) = 1.17 \times (-2.156 + 1.509/\lambda - 0.198/\lambda^2 + 0.011/\lambda^3) + 1.78 \\ {\rm for} \ 0.12\  \rm{\mu m} \leq \lambda < 0.63\  \rm{\mu m}
\end{multline}

Therefore, the SFR of each pixel is estimated from the extinction-corrected H$_{\alpha}$ luminosity as presented by \citet{1998ARA&A..36..189K} with the Salpeter initial mass function \citep[IMF,][]{1955ApJ...121..161S}:
\begin{equation}
  {\rm SFR} (M_{\odot}yr^{-1}) = 7.9 \times 10^{-42} L_{\rm H_{\alpha}}
\end{equation}

The mixing between gas and dust assumed in the extinction law of \citet{2001PASP..113.1449C} may not apply to the nearby S0s, as it has been found that the emission gas of some S0s is mainly located on their disk structures \citep[][]{2022A&A...659A..46B}. Using this method to correct the observational data may overestimate star formation in the galaxy center regions, but this overestimation does not depend on the environment. In this paper, when calculating the quantities related to SF (e.g., SFR), we performed spatially resolved BPT diagnosis (similar to Figure \ref{global-bpt}) for each galaxy, concentrating on the star-forming pixels. Similarly, we also require that the S/N of the emission lines ($H_{\alpha}$ $\lambda6563$, $H_{\beta}$ $\lambda4861$, [O\,{\sc ii}]$\lambda$$\lambda3726, 3729$, [O\,{\sc iii}]$\lambda$$\lambda4959, 5007$, and [N\,{\sc ii}] $\lambda6583$) $\geq$ 3.

\subsection{Radial profiles} \label{radial profiles}

\begin{figure*}[ht]
    \centering
    \includegraphics[scale=0.50]{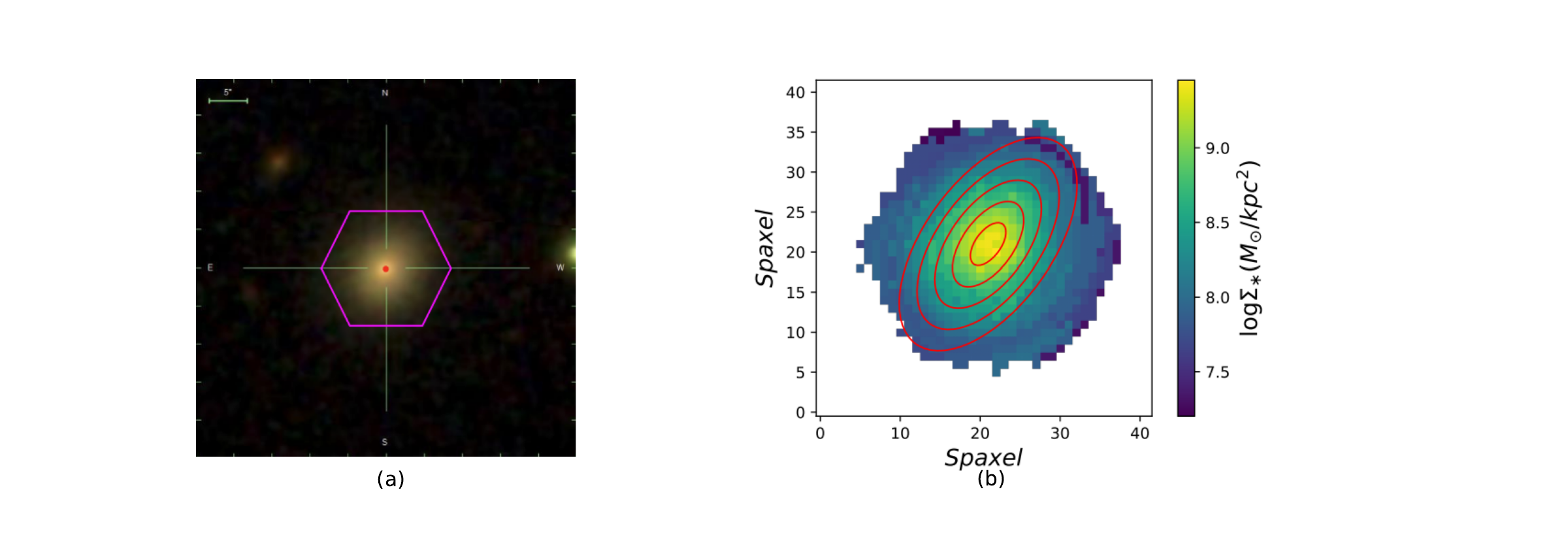}
    \caption{The SDSS optical image of the target and the radial profile constructed on the stellar mass density map. a)  SDSS $g, r, i$ color image. b) Aperture drawn over the $\Sigma_{\ast}$ map of galaxy PLATEIFU 11753-3701 in our sample. The ellipticity ($\varepsilon$) and position angle (PA) of the apertures are the same as those of this galaxy. The color is the value of each pixel and the color bar is the logarithmic display.}
    \label{11753-3701}
\end{figure*}

Quantities such as SFR surface density ($\Sigma_{\rm SFR}$) and resolved sSFR (rsSFR) are crucial for studying galaxy evolution at spatially resolved scales using IFS data \citep[e.g.,][]{2018MNRAS.474.2039E, 2020MNRAS.499..230B}. Analyzing physical quantities in galaxies through radial profiles, which represent the azimuthally averaged quantities as a function of distance from the galaxy's center, is a convenient approach. To compute these radial profiles, we utilize the Astropy-affiliated package Photutils: ``photutils.aperture" \citep[][]{2022zndo...7419741B}. The construction of our radial profiles is detailed as follows:

\begin{itemize}
    \item We specify the surface brightness peak of the galaxy as the center for constructing the aperture. To ensure alignment with the galaxy's center and major axis, we visually inspected the deeper DESI images of each galaxy. Briefly, elliptical apertures were created on the maps of $\Sigma_{\rm SFR}$, rsSFR, D$_{\rm n}$4000, and gas-phase metallicity. The ellipticity ($\varepsilon$) and position angle of each galaxy determine the parameters for constructing these apertures. Subsequently, we obtain the rsSFR map by taking the ratio of $\Sigma_{\rm SFR}$ and $\Sigma_{\ast}$ maps pixel-by-pixel.
    
    \item The semi-major axis of the central elliptical aperture was established at 0.2R$_{\rm e}$, followed by the creation of successive annuli with a thickness of 0.2R$_{\rm e}$ (0.2R$_{\rm e}$ bin) along the major axis of the galaxy. This process continued until the semi-major axis of the outermost ellipse reaches the boundary of the galaxy's IFU field of view. Figure \ref{11753-3701} is an example of our construction on the $\Sigma_{\ast}$ map of one of the galaxies. 
    
    \item Lastly, the median value represents the corresponding radial bins. For each galaxy, we utilized the observational error\footnote{In the MaNGA DAP map file, the inverse variance (Ivar) corresponding to each emission line is provided, which is equal to the reciprocal of the square of the corresponding physical quantity error, i.e., Error = 1/$\sqrt{\rm ivar}$} of H$_{\alpha}$ and the error of $\Sigma_{\ast}$ to compute the error of $\Sigma_{\rm SFR}$/rsSFR within a radial bin. Similarly, the observed errors of the emission lines ([N\,{\sc ii}] $\lambda6583$; [O\,{\sc ii}]$\lambda$$\lambda3726, 3729$; [O\,{\sc iii}]$\lambda$$\lambda4959, 5007$) and the continuous spectrum in the DAP map files were used to calculate the corresponding errors for gas-phase metallicity and D$_{\rm n}$4000, respectively.
\end{itemize}

Galaxies have a certain inclination angle, impacting surface density-related quantities \citep[e.g.,][]{2019ApJ...877..132W}. To correct the projection effect, a simple trigonometric correction involves multiplying by the scale factor ``$\rm sec(i)$", where ``i" is the inclination angle of the galaxy. The ``i" is defined as \citep{2022MNRAS.513..389R}:

\begin{equation}
    cos^{2}i = 
    \begin{cases}
    \frac{(1 - \varepsilon)^2 - (1 - \varepsilon_{\rm max})^2}{1 - (1 - \varepsilon_{\rm max})^2}, & \mbox{if }\varepsilon < \varepsilon_{\rm max} \\
    0, & \mbox{if }\varepsilon > \varepsilon_{\rm max}
    \end{cases}
    \label{de-projected}
\end{equation}
where $\varepsilon_{\rm max}$ is the approximate ellipticity exhibited by an edge-on SG, set at 0.8. The underlying assumption is that the outermost isophote of disk galaxies is approximately circular, with the inclination angle affecting only lengths perpendicular to the major axis. To ensure the validity of de-projection formulation (formula \ref{de-projected}), it is crucial that the inclination angle of the galaxy is below 75$^{\circ}$ \citep{2022MNRAS.513..389R}. By applying the definition of the ``i" (formula \ref{de-projected}) and $\varepsilon$\footnote{If the inclination angle of a galaxy is less than 75$^{\circ}$, the corresponding ellipticity is less than 0.68. According to the relationship between the ellipticity and axis ratio: $\varepsilon$ = 1 $-$ b/a, we require the axis ratio to be greater than 0.32 in Section \ref{sec2.1}.}, we identify the selected targets meeting the criterion, where the b/a of a galaxy is greater than 0.32. 

\subsection{Gas-phase metallicity}\label{section3.3}

The gas-phase metallicity serves as a crucial diagnostic tool as it tracks the immediate enrichment history of the interstellar medium (ISM) due to the stellar evolution and metal production across the galaxy. Since stellar evolution operates on longer timescales, short-term fluctuations in the gas outflow and inflow significantly impact the observed metallicity distributions. Therefore, studying gas-phase metallicity becomes a key approach to unraveling the origin and motions of gas on galactic scales \citep{2022A&A...659A.125S}. The two primary formation pathways of ``faded spiral" and ``minor merger" are intricately linked to the interaction types between the galaxy and its environments \citep[e.g.,][]{2013pss6.book..207V, 2019ApJ...872..144H, 2022A&A...659A..46B}. A significant distinction between these two approaches lies in whether the galaxy acquires new gas (``minor merger") or loses its original gas (``faded spiral"). \citet{1991ApJ...371...82S} showed that some SGs located in the Virgo cluster exhibit an increase in the metallicity of nearly 0.2 dex, while galaxies located on the outskirts of the Virgo cluster have similar abundances, compared to their counterparts in sparse environments. Of course, \citet{2006PASP..118..517B} used a larger sample to show that gas-poor LTGs in nearby clusters are more metal-rich than field galaxies. All these studies suggest that gas content may affect the chemical abundance of galaxies and that H\,{\sc i} deficiency often occurs when galaxies are forced to lose gas due to environmental effects, which may indicate that the environment of the galaxy played a role in the chemical evolution of the galaxy \citep[e.g.,][]{1991ApJ...371...82S, 2006PASP..118..517B, 2013A&A...550A.115H}. Considering the fact that the external fresh gas \citep[circumgalactic medium, CGM; intergalactic medium, IGM; companions or satellite;][]{2002ApJS..142...35K, 2019ApJ...872..144H} is mainly at a low metallicity \citep{2019ApJ...872..144H, 2022A&A...659A.125S}, the gas-phase metallicity gradient in a system with external fresh gas merging or inflow should be either flat or diluted \citep[e.g.,][]{2008AJ....135.1877E, 2010ApJ...721L..48K}. Therefore, we attempted to investigate the differences in the gas-phase metallicity of SFS0s in different environments, as described in Sect. \ref{Results}.


\section{Results}
\label{Results}

\subsection{Global properties}\label{global properties}

\begin{figure*}[ht]
    \centering
    \subfloat[\label{sersic}]{\includegraphics[width=0.45\textwidth]{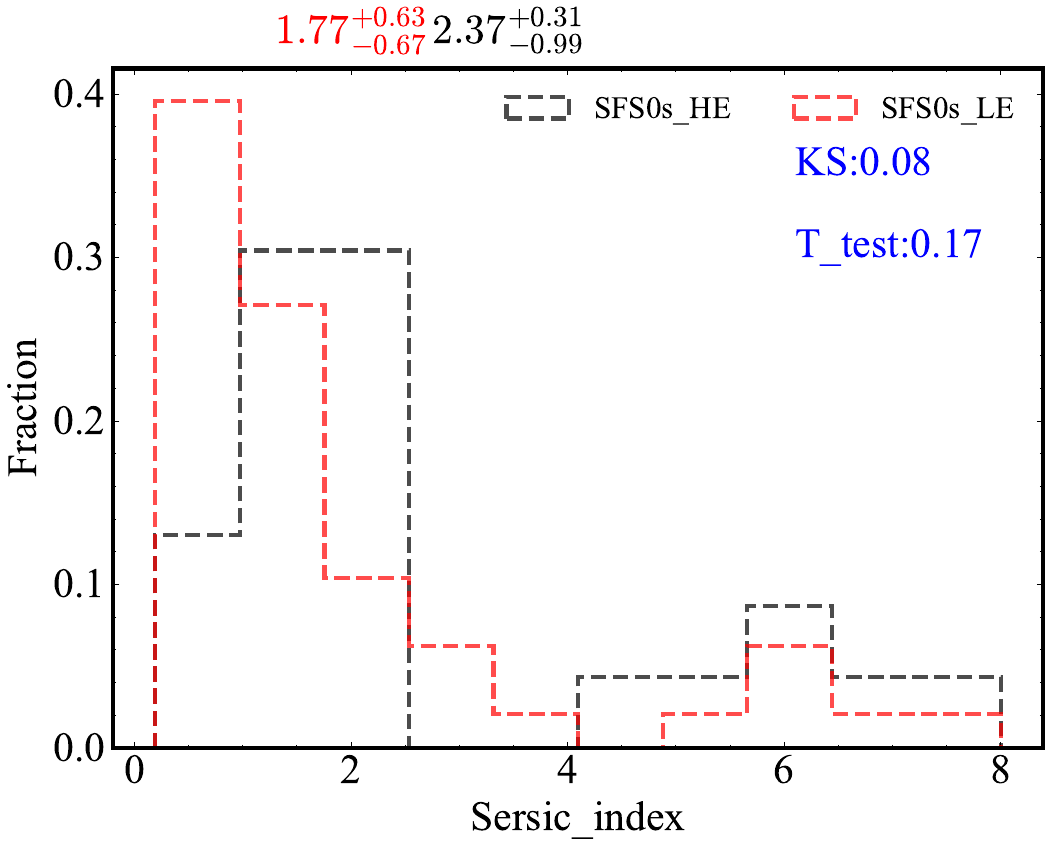}}
    \hspace{0.05\textwidth}
    \subfloat[\label{stellar}]{\includegraphics[width=0.45\textwidth]{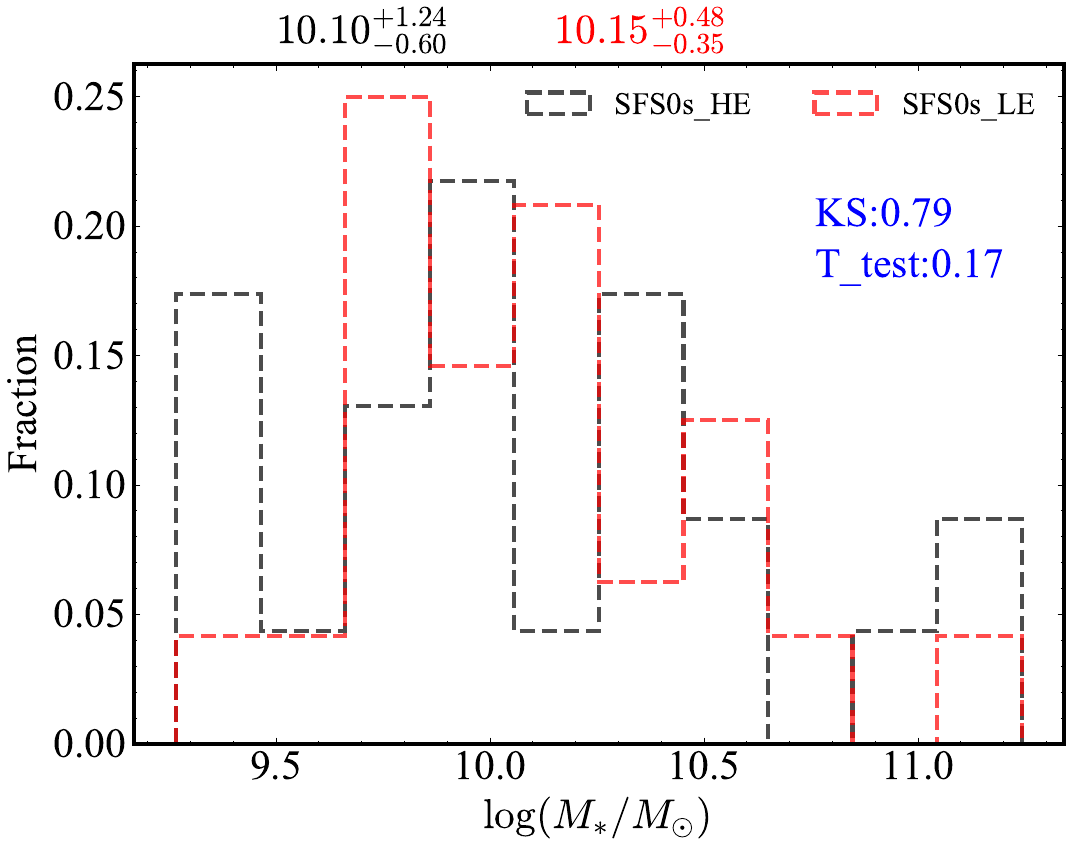}} \\
    \vspace{0.005\textwidth}
    \subfloat[\label{sm relation}]{\includegraphics[width=0.45\textwidth]{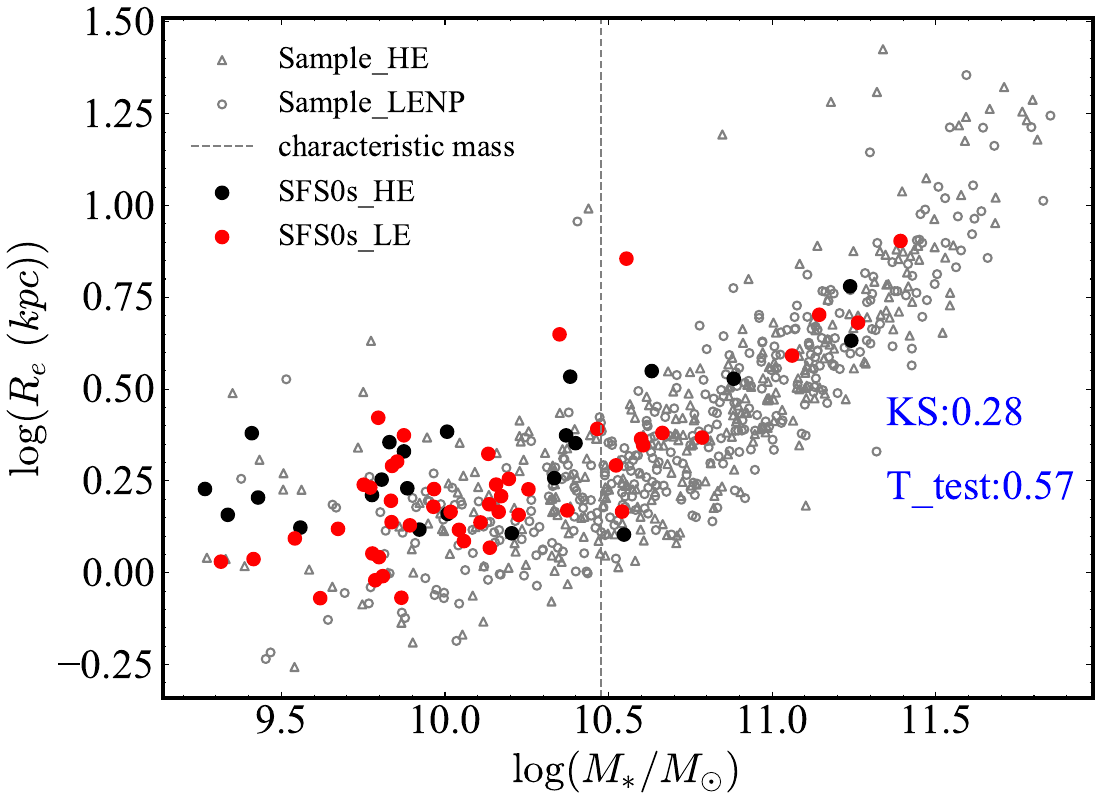}}
    \hspace{0.05\textwidth}
    \subfloat[\label{BT distribution}]{\includegraphics[width=0.45\textwidth]{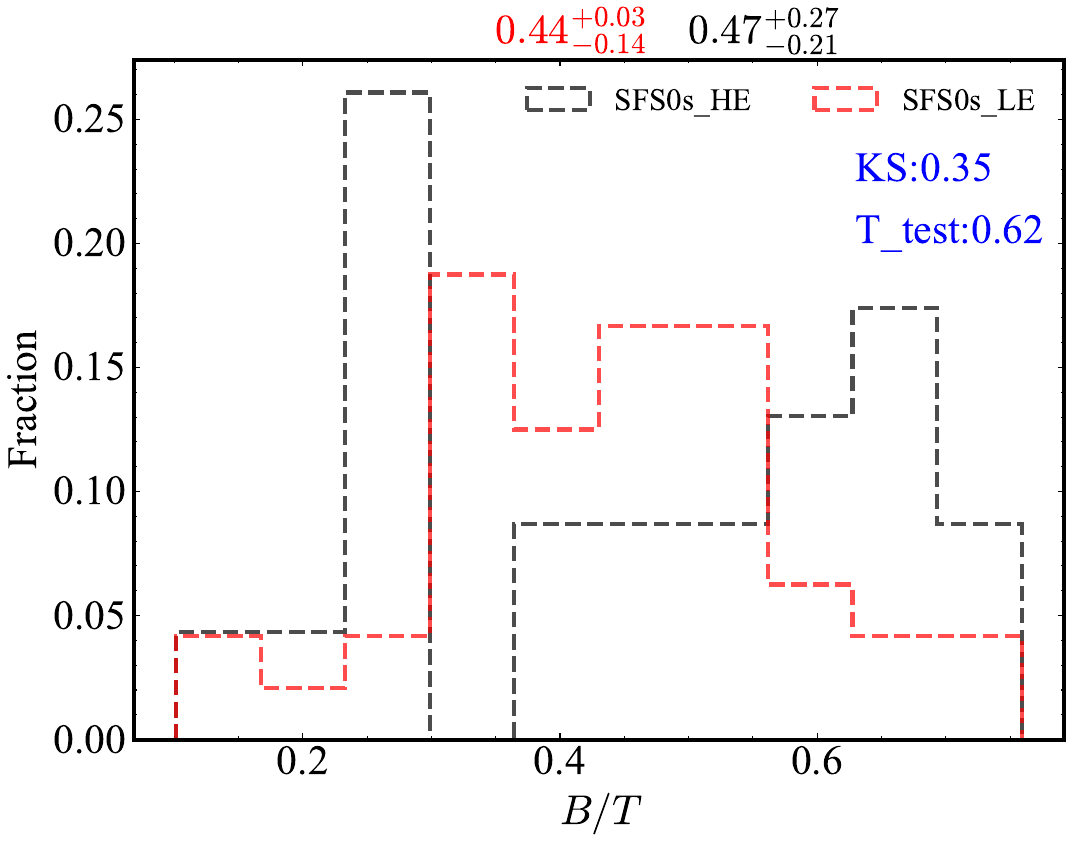}}
    \caption{The global properties of our two samples. a)  Distributions of the bulges S$\acute{\rm e}$rsic index, which come from MaNGA$\_$VAC photometrical catalog; b) Stellar mass (M$_{\ast}$) distribution; c) SMR. The gray dashed line represents the characteristic mass (``pivot mass", M$_{\ast} \approx 3 \times 10^{10}$) of the SMR bend \citep[][]{2016ARA&A..54..597C, 2019ApJ...872L..13M, 2021ApJ...921...38K}; In this figure, black and red circles represent our two samples and the rest are the same as Fig. \ref{global-bpt}; d) The bulge-to-total (B/T) light ratio distribution from Ser+Exp fit, which also comes from the photometrical catalog (r-band). In this picture, the top numbers represent the average of the parameters for our two samples, while the blue numbers in the picture represent different test results. The color markings are the same as in Fig. \ref{global-bpt}.}
    \label{Sersic_stellar distribution}
\end{figure*}

The different environments (i.e., formation pathways) will leave characteristic signatures in the newly formed S0s \citep[e.g.,][]{2022MNRAS.515..201C}. Figure \ref{sersic} illustrates the distribution of the bulge S$\acute{\rm e}$rsic index derived from the Ser+Exp fit \citep[][]{2022MNRAS.509.4024D} in the r-band for our samples. The S$\acute{\rm e}$rsic index values across our samples range from 0 to 8 \citep[][]{2009MNRAS.393.1531G, 2016ApJ...831...63X, 2022MNRAS.515..201C, 2022MNRAS.509.1237X}, with mean values of 2.37$^{+0.31}_{-0.99}$ for SFS0s$\_$HE and 1.77$^{+0.63}_{-0.67}$ for SFS0s$\_$LE. To compare the difference of S$\acute{\rm e}$rsic index distribution and the mean value of the two samples, we performed Kolmogorov-Smirnov test (KS$\_$test) and Student's t-test (hereafter T$\_$test) respectively (Table \ref{tab: parameter}). The test results show that the two distributions are identical (KS$\_$test: 0.08), and their mean values are also statistically consistent (T$\_$test: 0.17). For the overall SFS0s (mean: 1.96$^{+1.11}_{-0.85}$), the galaxies indeed demonstrate a higher incidence of pseudo bulges, consistent with the previous studies \citep[e.g.,][]{2022MNRAS.509.1237X}. If we truncate with S$\acute{\rm e}$rsic index = 2, we find that over 70$\%$ of galaxies in SFS0s$\_$LE have S$\acute{\rm e}$rsic index $<$ 2. However, the ratios of SFS0s$\_$HE are close (Table \ref{tab: composition}). 

\citet{2022MNRAS.513..389R} observed a pronounced reduction in the number of SFS0s beyond a stellar mass (M$_{\ast}$) threshold of 10.25 (log(M$_{\ast}$/$M_{\odot}$)), suggesting a potential manifestation of mass-deriven quenching. However, as they found, the relationship between mass quenching and SFS0s is unclear and requires further study of M$_{\ast}$ and morphology, especially based on IFS data; however, we know that mass quenching does not depend on the galaxy's environment \citep[e.g.,][]{2022A&A...666A.141M}. In Fig. \ref{stellar}, we present the M$_{\ast}$ histogram, with mean values of 10.10$^{+1.24}_{-0.60}$ (SFS0s$\_$HE) and 10.15 $^{+0.48}_{-0.35}$ (SFS0s$\_$LE). The KS$\_$test and T$\_$test both show that there is no difference between them (KS$\_$test: 0.79; T$\_$test: 0.17). For this reason, we do not discuss the effect of M$_{\ast}$ in the subsequent analysis.

\begin{table}
\footnotesize
\caption{Global and derivative properties of our samples\label{tab: parameter}}
\centering 
\begin{tabular}{ccccc} 
\hline\hline 
Parameter & SFS0s$\_$LE & SFS0s$\_$HE & KS$\_$test & T$\_$test\\
(1) & (2) & (3) & (4) & (5)\\
\hline 
S$\acute{\rm e}$rsic index & 1.77$^{+0.63}_{-0.67}$ & 2.37$^{+0.31}_{-0.99}$ & 0.08 & 0.17 \\
logM$_{\ast}/M_{\odot}$ & 10.15$^{+0.48}_{-0.35}$ & 10.10$^{+1.24}_{-0.60}$ & 0.79 & 0.17 \\
R$_{\rm e}$ (kpc) & 2.03$^{+0.93}_{-0.74}$ & 2.53$^{+1.68}_{-0.54}$ & 0.28 & 0.57 \\
B/T & 0.44$^{+0.03}_{-0.14}$ & 0.47$^{+0.27}_{-0.21}$ & 0.35 & 0.62 \\
$CI_{\rm H_{\alpha}/cont}$ & 0.07$^{+0.10}_{-0.25}$ & $-0.04^{+0.27}_{-0.38}$ & $5 \times 10^{-3}$ & 0.01 \\
$CI_{\rm SFR, H_{\alpha}}$ & 2.34$^{+0.22}_{-0.33}$ & $1.99^{+0.09}_{-0.19}$ & $2 \times 10^{-7}$ & $2 \times 10^{-3}$ \\
\hline 
\end{tabular}
\\
Notes: In this table, we represent each parameter as an average value with its FWHD of the distribution.
\end{table}

Furthermore, the size-mass relation (SMR) of galaxies serves as a valuable tool for exploring galaxy evolution \citep[][]{2013MNRAS.432.1862C, 2016ARA&A..54..597C}. Figure \ref{sm relation} displays the SMR, with R$_{\rm e}$ (r-band) measured in ``kpc." We find a bend in this relation at the low-mass end, which may be due to a break in the SMR at M$_{\ast}$ $\sim$ 3 $\times$ 10$^{10}$, known as a characteristic mass \citep[``pivot mass;" e.g.,][]{2013MNRAS.432.1862C, 2016ARA&A..54..597C, 2019ApJ...872L..13M, 2021ApJ...921...38K}. This mass is considered to have physical significance as it is closer to the M$_{\ast}$ threshold, beyond which 50$\%$ of galaxies are quenched \citep[e.g.,][]{2019ApJ...872L..13M, 2021ApJ...921...38K}. Moreover, \citet{2013MNRAS.432.1862C} suggested that this bending corresponds to the two main formation pathways of ETGs. In Fig. \ref{sm relation}, we find that SFS0s$\_$LE, on average, has slightly smaller R$_{\rm e}$ values (mean: 2.03$^{+0.93}_{-0.74}$ kpc; 2.53$^{+1.68}_{-0.54}$ kpc for SFS0s$\_$HE). In other words, SFS0s$\_$LE galaxies are more compact for a given M$_{\ast}$, although both tests show that there is no statistical difference in R$_{\rm e}$ (see Table \ref{tab: parameter}). Table \ref{tab: composition} provides a subdivision of our samples based on the M$_{\ast}$ criterion from \citet{2022MNRAS.513..389R}, presenting the proportions of high/low mass galaxies in our samples. Their proportions are also comparable. Figure \ref{BT distribution} displays the histogram of the bulge-to-total light ratio (B/T). We find that the mean value of SFS0s$\_$LE is 0.44$^{+0.03}_{-0.14}$ and SFS0s$\_$HE is  0.47$^{+0.27}_{-0.21}$. The KS$\_$test and T$\_$test both indicate that there is no difference between them (KS$\_$test: 0.35; T$\_$test: 0.62). When considering our all samples (mean: 0.45$^{+0.17}_{-0.11}$), this result is aligned with previous works.

\subsection{SFR CI}\label{CBCI_para}

\begin{figure}
    \centering
    \includegraphics[width=\linewidth]{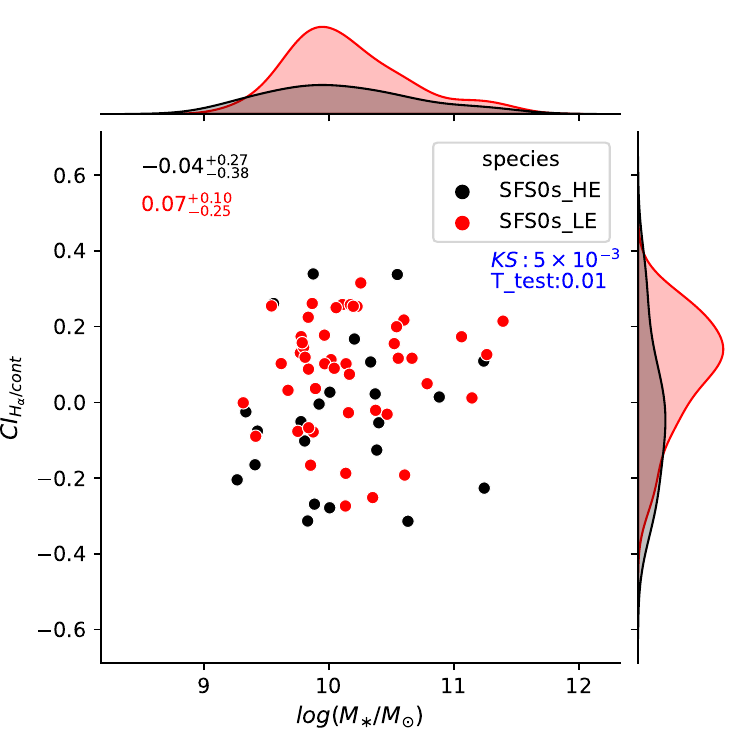}
    \caption{$CI_{\rm H_{\alpha}/cont}$ values based on the concentration defined (Formula \ref{ci_wang}) from \citet{2022MNRAS.516.3411W}
(black: SFS0s$\_$HE; red: SFS0s$\_$LE). The mean $CI_{\rm H_{\alpha}/cont}$ of galaxies in HDEs and LDEs is $-0.04^{+0.27}_{-0.38}$ and $0.07^{+0.10}_{-0.25}$, respectively. The curves on the top and right illustrate the distributions of the parameters on the horizontal and vertical coordinates. The KS$\_$test and T$\_$test results are represented by the blue numbers.}
    \label{CI_WD}
\end{figure}

Previous studies have proposed various morphological parameters to distinguish between ETGs and LTGs \citep[e.g.,][]{2009A&A...497..743H, 2019MNRAS.485..452M, 2020MNRAS.491.3624B}, where the concentration index ($CI$) describes the distribution of light among the associated pixels within a galaxy \citep[][]{1996ApJS..107....1A, 2000AJ....119.2645B, 2003ApJ...588..218A, 2004AJ....128..163L}. \citet{2022MNRAS.516.3411W} used the concentrated SF and stellar population (SP) ages to study the effects of environmental quenching on all galaxies in the SAMI Survey. They defined their $CI$ parameter ($CI_{\rm H_{\alpha}/cont}$) of the gas disk relative to the stellar disk:

\begin{equation}\label{ci_wang}
    CI_{\rm H_{\alpha}/cont} = {\rm log}(r_{\rm 50,H_{\alpha}}/r_{\rm 50,cont})
\end{equation}
where ``cont" refers to the $r$-band continuum of their galaxies and $r_{50}$ represents the radius containing 50$\%$ of the emission line or continuum luminosity. In Figure \ref{CI_WD}, we give the distributions of $CI_{\rm H_{\alpha}/cont}$, where blue numbers represent the test (KS$\_$test and T$\_$test) of $CI_{\rm H_{\alpha}/cont}$ distributions and mean values, and black and red numbers are the mean values of $CI_{\rm H_{\alpha}/cont}$. We find that the $CI_{\rm H_{\alpha}/cont}$ is larger ($0.07^{+0.10}_{-0.25}$) in SFS0s$\_$LE and smaller in SFS0s$\_$HE ($-0.04^{+0.27}_{-0.38}$). The KS$\_$test shows that the difference of the $CI_{\rm H_{\alpha}/cont}$ parameter is statistically significant (KS$\_$test: $5 \times 10^{-3}$), and their means are also inconsistent (T$\_$test: 0.01). In other words, the distributions of gas in SFS0s on the gas disk are different in different environments. Of course, the variation in the gas disk present in the two samples relative to the stellar disk can be called the shrinking effect of the gas disk. Previous findings indicate that the H\,{\sc i} disk of a galaxy responds to rem-pressure stripping (RPS) by shrinking \citep[][]{2023ApJ...956..148L}, while the stellar disk remains unaffected.

\begin{figure*}[ht]
    \centering
    \subfloat[]{\includegraphics[width=0.45\textwidth]{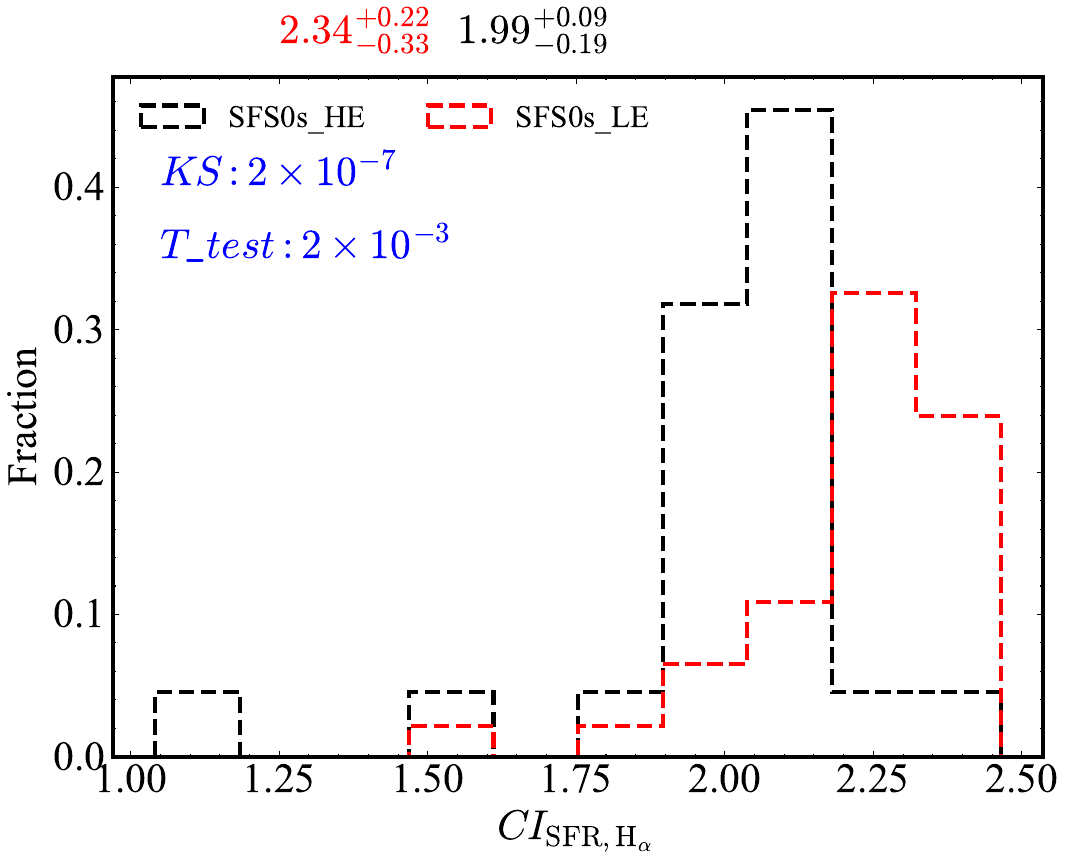}\label{CBCI_dis}}
    \hspace{0.05\textwidth}
    \subfloat[]{\includegraphics[width=0.45\textwidth]{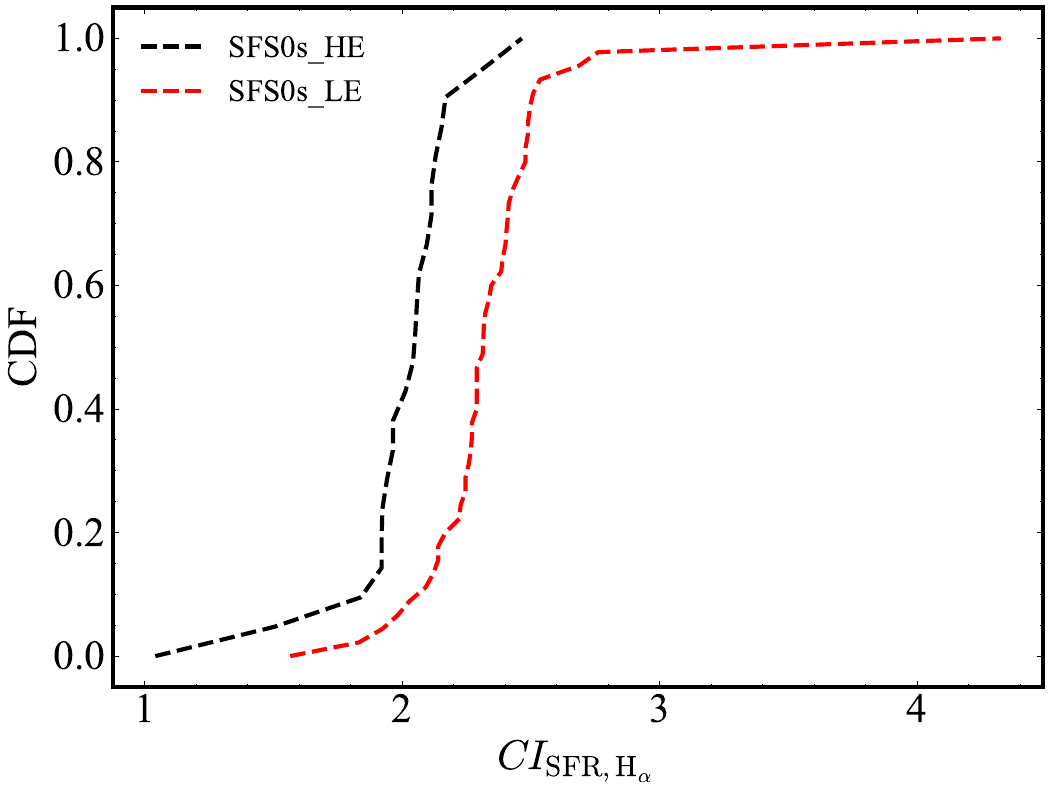}\label{CBCI_CDF}}
    \caption{$CI_{\rm SFR, H_{\alpha}}$ distributions and its Cumulative distribution function (CDF). a) $CI_{\rm SFR, H_{\alpha}}$ distributions. The blue numbers represent the results of the KS$\_$test and T$\_$test, and the different color numbers are the mean value of the two samples; b) The CDF of the $CI_{\rm SFR, H_{\alpha}}$. Color markings are the same as in Figure \ref{global-bpt}.} 
    \label{CBCI}
\end{figure*}

The two origin mechanisms proposed above make the disturbance of the gas and stars of the galaxy different in different environments \citep[e.g.,][]{2020MNRAS.498.2372D, 2021MNRAS.508..895D, 2022A&A...659A..46B}. Thus, we only focus on the gas disk of the galaxy to prove the difference in $CI_{\rm H_{\alpha}/cont}$ parameter. \citet{2000AJ....119.2645B} provided the concentration index ($CI$), which is defined as the logarithm of the ratio of the circular radii containing 80$\%$ and 20$\%$ of the total flux \citep[][]{2000AJ....119.2645B}:

\begin{equation}\label{ci f}
    CI = 5 \times {\rm log}(R_{80}/R_{20})
\end{equation}
Typically, higher $CI$ values correspond to higher fractions of light in the central regions, suggesting that the galaxy is less disturbed \citep[][]{2022MNRAS.tmp..901G}. In the case of our SFS0s, we apply the $CI$ (formula \ref{ci f}) to analyze the SFR of these galaxies ($CI_{\rm SFR, H_{\alpha}}$). Specifically, we define this parameter as the logarithm of the ratio of the circular radii containing 80$\%$ and 20$\%$ of the total SFR of star-forming pixels within the galaxy:

\begin{equation}\label{my ci}
    CI_{\rm SFR, H_{\alpha}} = 5 \times {\rm log}(R_{\rm 80,SFR}/R_{\rm 20,SFR})
\end{equation}
The $CI$ parameter (formula \ref{ci f}) is originally associated with the stellar continuum, but in this context, we adapt it (formula \ref{my ci}) to assess the distribution of ionized gas in galaxies, specifically the distribution of H$_{\alpha}$ emission. This adaptation assumes elliptical or circular distributions of H$_{\alpha}$. By examining the H$_{\alpha}$ emission, we find that the fraction of galaxies in our sample that exhibit significant clumps is negligible. Moreover, galaxies with star-forming rings \citep[e.g.,][]{2022A&A...659A..46B, 2023ApJ...942...48T} are only a few percent of our sample, and therefore do not have a noticeable impact on the analysis of $CI_{\rm SFR, H_{\alpha}}$, even though such ring structures are quite common in SFS0s \citep[][]{2022A&A...659A..46B}. The distributions of $CI_{\rm SFR, H_{\alpha}}$ are depicted in Figure \ref{CBCI_dis}, while the Cumulative distribution functions (CDF) are illustrated in Figure \ref{CBCI_CDF}. The color markings in both figures correspond to those in Figure \ref{global-bpt}. We find that the $CI_{\rm SFR, H_{\alpha}}$ parameter is larger in SFS0s$\_$LE (mean: $2.34^{+0.22}_{-0.33}$) and smaller in SFS0s$\_$HE ($1.99^{+0.09}_{-0.19}$). We provide the results of KS$\_$test and T$\_$test (blue numbers) for $CI_{\rm SFR, H_{\alpha}}$ parameter in this figure. We find that the $CI_{\rm SFR, H_{\alpha}}$ difference between the two samples is statistically significant (KS$\_$test: $2 \times 10^{-7}$), and their means also show significant differences (T$\_$test: $2 \times 10^{-3}$). Also, the CDF can tell us the same result (Figure \ref{CBCI_CDF}).

\begin{figure*}
    \centering
    \subfloat[]{\includegraphics[width=0.45\textwidth]{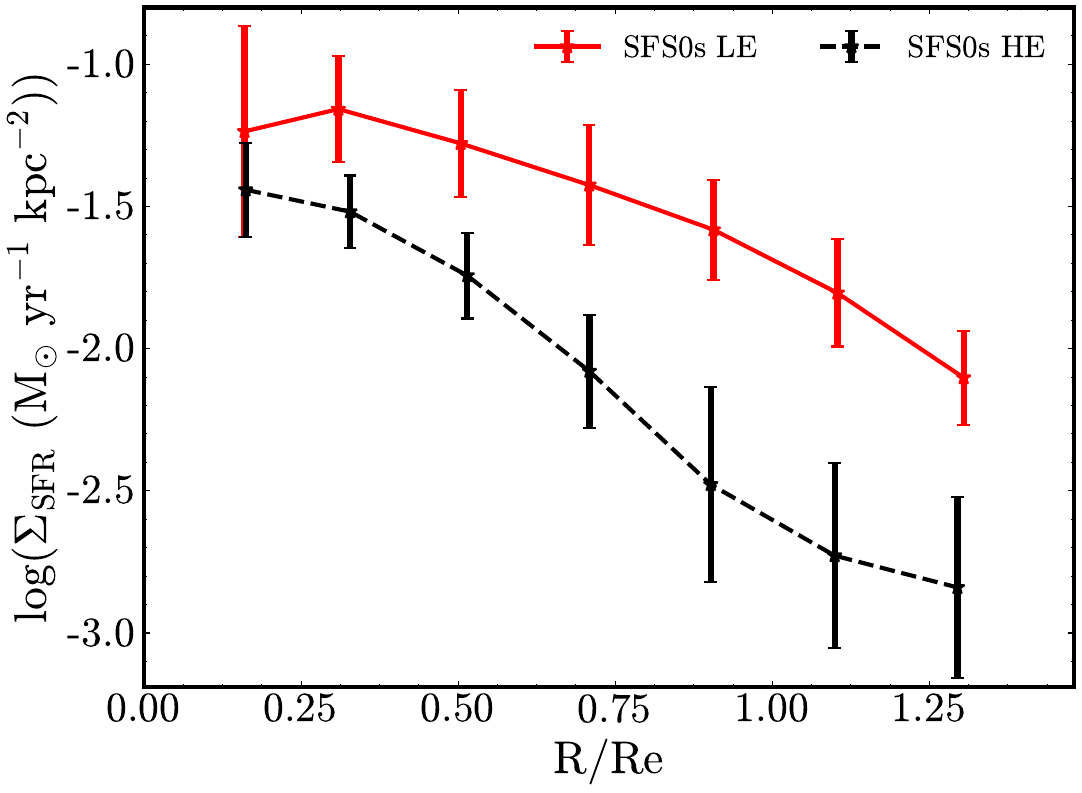}\label{sigma-SFR}}
    \hspace{0.05\textwidth}
    \subfloat[]{\includegraphics[width=0.45\textwidth]{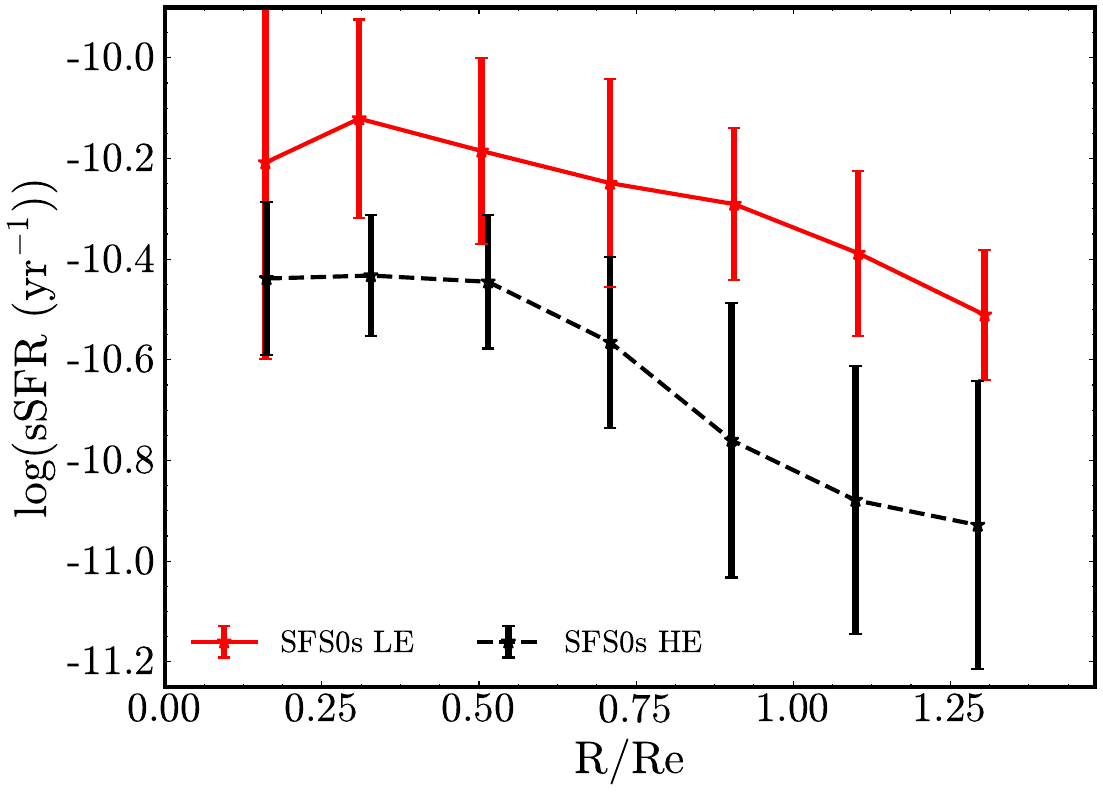}\label{sSFR-rp}}
    \caption{The radial profiles of $\Sigma_{\rm SFR}$ and rsSFR. a) $\Sigma_{\rm SFR}$ radial profiles; b) rsSFR radial profiles. Color markings of our samples are the same as in Figure \ref{global-bpt}. The error bars represent the standard deviation.}
    \label{sigma_SFR-sSFR}
\end{figure*}

\subsection{Radial profiles of derived quantities}\label{DQ radial profiles}

n Fig. \ref{sigma-SFR}, a noticeable central peak of $\Sigma_{\rm SFR}$ in both samples is observed, gradually diminishing with increasing radial distance. To depict the efficiency of gas conversion into stars within a galaxy, we present radial profiles of rsSFR in Fig. \ref{sSFR-rp}. This pattern is aligned with previous findings, such as the study by \citet{2022MNRAS.513..389R}, where it was suggested that their primary SF activities predominantly occur in the inner regions (Sect. \ref{CBCI_para}). Moreover, this also indicates that potential contamination (about 8$\%$) from SGs are not expected to affect our results. We find that the radial profiles of $\Sigma_{\rm SFR}$ and rsSFR for the two samples are different and even when we consider the error bars, there is no overlap between the two quantities. For LDEs, both quantities are higher than HDEs, but there is a more pronounced steepness in HDEs.

\begin{figure*}
    \centering
    \subfloat[]{\includegraphics[width=0.45\textwidth]{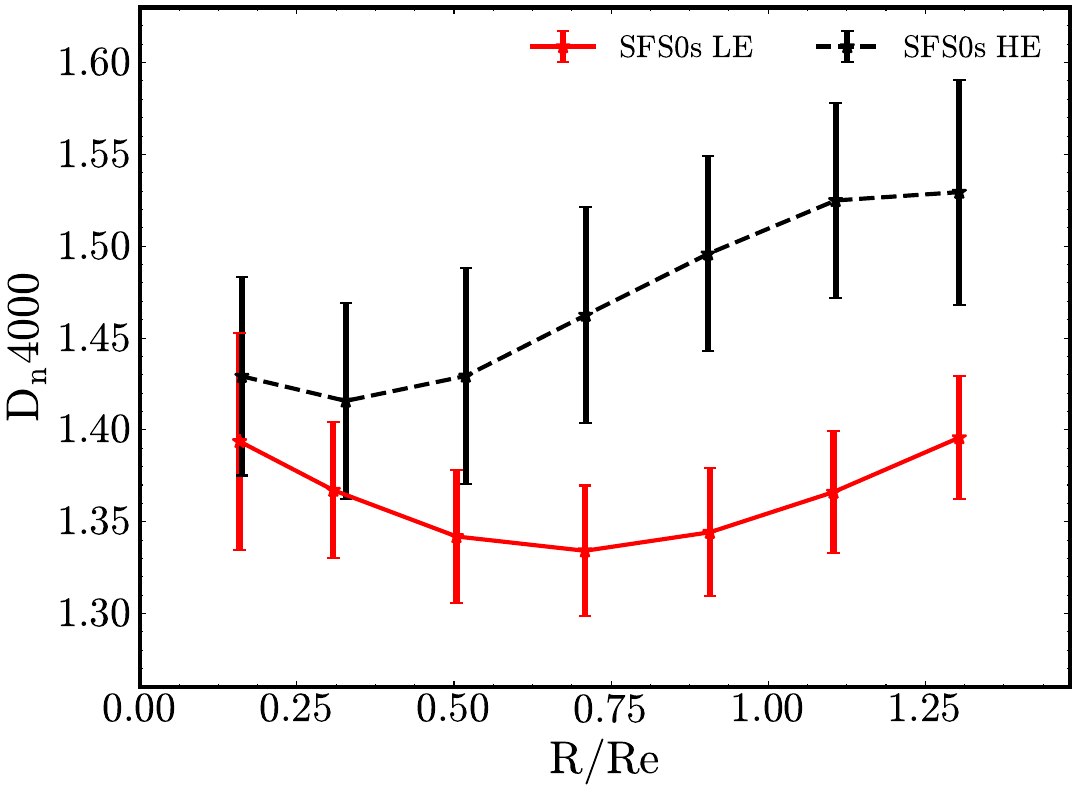}\label{dn4000_rp}}
    \hspace{0.05\textwidth}
    \subfloat[]{\includegraphics[width=0.45\textwidth]{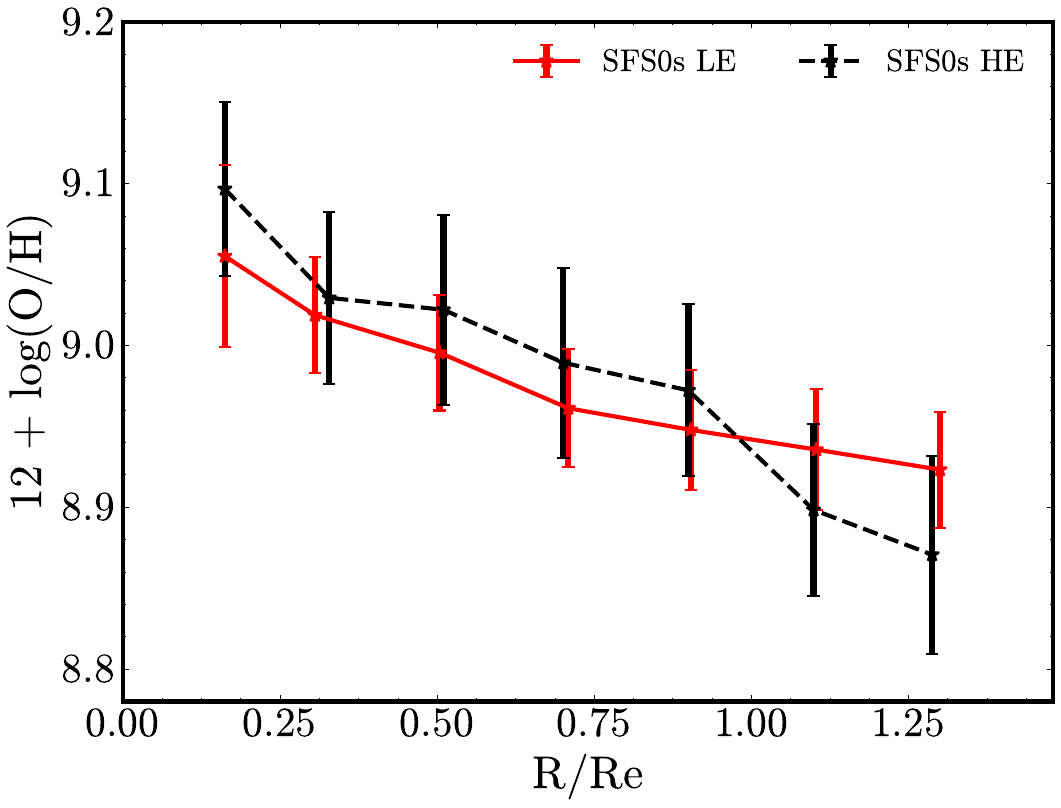}\label{meatllicity_rp}}
    \caption{The radial profiles of D$_{\rm n}$4000 and gas-phase metallicity. a) D$_{\rm n}$4000 radial profiles; b) Gas-phase metallicity radial profiles. The color markings are the same as in Figure \ref{global-bpt}. The error bars represent the standard deviation.}
    \label{dn4000-metallicity}
\end{figure*}

As mentioned above, different perturbations of SFS0s in different environments have different effects on the gas and star in the system ($CI_{\rm H_{\alpha}/cont}$; Fig. \ref{CI_WD}). This leads to a distinct difference in the distribution of SF in the galaxy, as traced by the concentration indices of SFR ($CI_{\rm SFR, H_{\alpha}}$; Figs. \ref{CI_WD} and \ref{CBCI_dis}). These characteristics are bound to affect the distribution of D$_{\rm n}$4000 in SFS0s. In Fig. \ref{dn4000_rp}, we provide the D$_{\rm n}$4000 radial profiles, which are the SP age gradients in different environments. We find that the SPs' age of the galaxy in HDEs increases with the increase of radial distance, which corresponds to the younger bulges and the older disk \citep[][]{2012MNRAS.422.2590J}; whereas it appears to be flatter in SFS0s$\_$LE. Furthermore, The SPs in LDEs are younger than that in HDEs.  

According to Sect. \ref{section3.3}, we employed the ([O\,{\sc ii}]$\lambda$$\lambda3726, 3729$ + [O\,{\sc iii}]$\lambda$$\lambda4959, 5007$)/$H_{\beta}$ (R$_{23}$, S/N of these lines $\geq$ 3) ratio as a tracer of the gas-phase metallicity \citep[][]{2004ApJ...613..898T, 2021MNRAS.505..191B} and the results are shown in Fig. \ref{meatllicity_rp}. The observed trend in our samples reveals a gradual decline in metallicity from the inner to outer regions, consistent with previous results \citep[e.g.,][]{2022NatAs...6.1464C, 2022A&A...659A.125S}. We find that the changing trends of the two are very similar and very closely aligned.  

\section{Discussion}\label{Discussion}

We find no statistical difference in the bulge S$\acute{\rm e}$rsic index, stellar mass, and B/T ratio among our two samples in Sect. \ref{global properties}. In addition, the T$\_$test also shows no difference in their mean values (Table \ref{tab: parameter}). Combined with the size of our sample and the fact that SFS0s only represent a very specific phase in the transition from gas-rich SGs to gas-poor S0s, we consider these results normal. When we consider these parameters of the total sample, it is consistent with previous works: contains more pseudo-bulges, smaller B/T ratios, and similar M$_{\ast}$ \citep[e.g.,][]{2016ApJ...831...63X, 2022MNRAS.512.2222V, 2022MNRAS.509.1237X}. We also find the bending on the SMR \citep[e.g.,][]{2013MNRAS.432.1862C, 2016ARA&A..54..597C}. In contrast, \citet{2022MNRAS.515..201C} combined kinematics, morphology, and properties of the stellar populations (SPs) of 329 S0s from SAMI/MaNGA to study the role of environment in the formation and evolution of S0s. These authors found that the sub-sample originating from the ``faded spiral" was (on average) younger, more rotationally supported, smaller, and with lower S$\acute{\rm e}$rsic index, less massive, and tended to have slightly smaller B/T values. However, the sub-sample originating from the ``merger" was older, more pressure supported, larger, and with larger S$\acute{\rm e}$rsic index, more massive, and tended to be slightly higher B/T. The difference between our results and \citet{2022MNRAS.515..201C} may be due to our sample is all S0s in the star-forming state (see Tables \ref{SFS0s_HE} and \ref{SFS0s_LE}) and it may be found that the distribution of SFR in our sample is different from theirs (refer to their Table 1).

We provide the $CI$ parameters ($CI_{\rm H_{\alpha}/cont}$ and $CI_{\rm SFR, H_{\alpha}}$) of the galaxy's SF in Sect. \ref{CBCI_para} to reflect the influence of the environment on the distribution of galaxy's SF. We find that the difference in $CI_{\rm H_{\alpha}/cont}$ between the two samples is significant (KS$\_$test: $5 \times 10^{-3}$; T$\_$test: 0.01) and can be attributed to different origin mechanisms in different environments. The origin of galaxies in SFS0s$\_$LE originates from: a minor merger, with the merger or inflow of fresh external gas \citep[e.g.,][]{2014MNRAS.438.2798K}, which leads to a more extended distribution of gas in these galaxies compared to stellar disks. However, galaxies in SFS0s$\_$HE experience gas stripping, losing gas from the outer disk \citep[RPS;][]{2013pss6.book..207V, 2022A&A...659A..46B}, or a blocking of the external gas supply \citep[starvation/strangulation;][]{2006ApJ...651..811B, 2016A&A...596A..11B, 2018A&A...614A..57F, 2022A&A...659A..46B}. This results in a more compact distribution of gas compared to stellar disks (see Fig. \ref{CI_WD}), which has been confirmed by the shrinking effect of the H\,{\sc i} disk reported in a previous work \citep[][]{2023ApJ...956..148L}. Therefore, SFS0s exhibits a larger $CI_{\rm H_{\alpha}/cont}$ ($0.07^{+0.10}_{-0.25}$) in SFS0s$\_$LE, while SFS0s$\_$HE exhibits a smaller $CI_{\rm H_{\alpha}/cont}$ ($-0.04^{+0.27}_{-0.38}$).

To confirm this discrepancy, we referred to the $CI$ parameters of light within galaxies and defined a $CI_{\rm SFR, H_{\alpha}}$ parameter that only considers the gas disk in the galaxy. Combining the meaning of $CI_{\rm H_{\alpha}/cont}$, the gas disk in SFS0s$\_$HE produces a shrinking effect \citep[][]{2023ApJ...956..148L}, and these galaxies either lose gas from the outer disk or have no external gas supply. Therefore, the SF of these galaxies is either truncated or uniformly stopped after consuming gas reservoir \citep[][]{2006ApJ...651..811B, 2016A&A...596A..11B, 2018A&A...614A..57F, 2022A&A...659A..46B}. However, the SFS0s$\_$LE can have a supply of external gas and even fall into the nuclear regions of the galaxy \citep[][]{2002ApJS..142...35K, 2019ApJ...872..144H}, so the SF of the outer regions for these galaxies may be high in comparison, depending on the angular momentum of the gas reaching the outer disk of the galaxy \citep[][]{2019ApJ...883L..36H}. Therefore, the galaxies in SFS0s$\_$LE exhibit higher $CI_{\rm SFR, H_{\alpha}}$ ($2.34^{+0.22}_{-0.33}$), while SFS0s$\_$HE shows lower parameter ($1.99^{+0.09}_{-0.19}$). The difference is statistically significant (KS$\_$test: $2 \times 10^{-7}$ and T$\_$test: $2 \times 10^{-3}$). By comparing the $R_{\rm 80, SFR}$ (SFS0s$\_$LE median: 3.00 kpc; SFS0s$\_$HE: 2.48 kpc) and R$_{\rm e}$ of our galaxies with stellar mass smaller than the ``pivot mass," we are able to obtain similar results. The consistency between the $R_{\rm 20, SFR}$ (SFS0s$\_$LE median: 0.76 kpc; SFS0s$\_$HE: 0.59 kpc; KS$\_$test: 0.13) of the two samples also seems to indicate that the SF of these SFS0s is center-dominated rather than disk-dominated \citep[][]{2022MNRAS.513..389R}. It is crucial to emphasize that our application of $CI_{\rm SFR, H_{\alpha}}$ only considers the gas disk of the galaxy, excluding the stellar disk. Certainly, further validation through observations or simulations that demonstrate changes in the gas gradient on the gas disks of galaxies subjected to perturbed processes would enhance the credibility of our findings. However, it is disappointing that no consensus has been reached so far. Ongoing advancements in observational techniques and simulation capabilities could potentially shed more light on these aspects, contributing to a deeper understanding of the relationship between environments, perturbation processes, and gas distribution in galaxies.

Here, we find that there are obvious differences in the SF distributions of SFS0s in different environments, so we further study the radial profile of the galaxy's derivation quantities. In Fig. \ref{sigma_SFR-sSFR}, we find that on average, SFS0s$\_$LE has a higher $\Sigma_{\rm SFR}$/rsSFR, which is attributed to the injection of fresh external gas. These newly entered gases may not immediately participate in SF, only igniting the SF of the galaxy when the gas loses its original angular momentum and cooling \citep[][]{2019ApJ...883L..36H}. This is consistent with the origin of the ``minor merger." However, the disturbance mechanisms of SFS0s$\_$HE either force the gas in the outer disk to be stripped or stop the supply of gas for the galaxy, so a low $\Sigma_{\rm SFR}$/rsSFR is understandable. This is consistent with a ``faded spiral" origin. Of course, to confirm this result requires testing the content of atomic or molecular gas in SFS0s (Chen et al in prep). In addition, the central peak $\Sigma_{\rm SFR}$ of both samples does indicate that their SF activities are concentrated in the galaxy's center \citep[][]{2022MNRAS.513..389R}.

It is precisely thanks to the different origin mechanisms that we have discovered differences in SP ages (D$_{\rm n}$4000; see Fig. \ref{dn4000_rp}). As mentioned earlier, the merger-like events experienced by galaxies in LDEs involve the merging or inflow of external gas \citep[e.g.,][]{2014MNRAS.438.2798K}, even forcing this gas to fall into the galaxy's nuclear regions \citep[][]{2002ApJS..142...35K, 2019ApJ...872..144H}. The dissipation of angular momentum \citep[][]{2019ApJ...883L..36H} and the cooling of the gas during the fall will ignite the SF of the corresponding regions, resulting in the formation of young massive stars in these regions; thus, their SP ages will be slightly younger (lower D$_{\rm n}$4000). For the SFS0s$\_$HE sample, however, the galaxy undergoes a perturbation process that either causes the galaxy to preferentially lose gas from the outer disk or prevents the supply of external gas, thereby demonstrating a gradually increasing D$_{\rm n}$4000 value \citep[e.g.,][]{2022ApJ...931L..22L, 2023MNRAS.523.1140L}. The change in the SP age gradient in SFS0s$\_$HE seems to correspond to an outside-in quenching mechanism \citep[][]{2018A&A...614A..57F, 2019ApJ...872...50L}. 

However, what was most surprising  was the fact that the origin mechanism of SFS0s$\_$LE requires the presence of external fresh gas merging or inflow. These metal-poor gases from CGM, IGM, companion or satellite galaxies \citep[][]{2002ApJS..142...35K, 2019ApJ...872..144H} will significantly dilute the gas phase metallicity of the galaxy \citep[][]{2008AJ....135.1877E, 2010ApJ...721L..48K}. In contrast, we did not find any significantly lower gas-phase metallicity in SFS0s$\_$LE (see Fig. \ref{meatllicity_rp}). After the gas in the outer disk of the galaxy is stripped away, it will leave behind obvious metal-rich systems \citep[][]{2013A&A...550A.115H}. However, we did not find any increased metallicity in SFS0s$\_$HE either. As described in Sect. \ref{global properties}, although the R$_{\rm e}$ of the two samples follows the same statistical distribution, we can observe that the galaxies in SFS0s$\_$LE have on average smaller sizes, especially those below the ``pivot mass," which leads to higher gravitational potential \citep[$\Phi$ $\propto$ M$_{\ast}/{\rm R}_{\rm e}$,][]{2022MNRAS.516.2971V}. Therefore, the feedback processes present in galaxies may not be sufficient to remove metals from such systems \citep[e.g.,][]{2020A&A...634A..95C}, leading to an increase in metallicity in SFS0s$\_$LE. In fact, the gas-phase metallicity is the result of competition between gas inflow, outflow, supernova feedback, and SF feedback. It is only when the inflow of metal-poor gas is strong and the SF does not last not long enough to enrich the interstellar medium (ISM) through supernova explosions, the galaxy may exhibit metallicity deficiency \citep[][]{2024ApJ...968....3H}. In the studies of \citet{2022NatAs...6.1464C} and \citet{2024ApJ...968....3H}, no significantly lower metallicity was found in systems that may have the external fresh gas supply. Of course, we checked SFS0s of different M$_{\ast}$ and excluded the influence of M$_{\ast}$, which gave the same result. Moreover, using the measurements from the central 3$^{\prime \prime}$ within galaxies provided by MPA-JHU \citep[][]{2004ApJ...613..898T}, we also observed the same result. Previous works have shown that gas-rich galaxies in rich environments have, on average, higher metallicity than their counterparts in sparse environments \citep[][]{1991ApJ...371...82S, 2013A&A...550A.115H}. Similarly, this requires testing the gas content in these galaxies (Chen et al in prep).

Of course, it is worth noting how frequently these disturbance mechanisms exist in different environments within the M$_{\ast}$ range and redshift range of our sample. Because studies have found that the merging fraction varies significantly with redshift and the M$_{\ast}$ of galaxies \citep[e.g.,][]{2011ApJ...742..103L}. \citet{2023MNRAS.522....1N} studied 1.3 million galaxies from the SDSS DR16 photometric catalog and presented the probability that each galaxy is a major or minor merger, splitting the classifications into merger stages (early, late, and post-coalescence). They found that the fraction of minor merger occurring in massive galaxies (logM$_{\ast}$ $\geq$ 10.40 $M_{\odot}$) is approximately 20$\%$ - 65$\%$ at redshift z $\sim$ 0.01-0.05, while the fraction of major merger occurring is approximately 10$\%$ - 30$\%$ (z = 0-0.05) when the M$_{\ast}$ is not considered. If we assume that all the galaxies in the sample SFS0s$\_$LE come from the merger, then this fraction (68$\%$, 48/71, or massive fraction 21$\%$, 15/71) is consistent. In addition, simulations \citep[][]{2019MNRAS.483.1042Y} have shown that galaxies affected by RPS are common in galaxy clusters with a redshift z $<$ 0.6, with approximately 30$\%$ of disk galaxies having gas compositions resembling comet shapes \citep[][]{2019A&A...631A.114B}. Indeed, if we assume that all galaxies in SFS0s$\_$HE come from RPS, then this frequency is also consistent (32$\%$, 23/71). However, it is important to note that due to the size and specificity of our sample, we cannot draw any firm conclusions, therefore these results ought to be taken with caution.

\section{Conclusions}\label{Conclusions}

We assembled a sample of 71 SFS0s, utilizing data from the SDSS-IV MaNGA survey and subsequently segregating it into two categories based on environmental density information: SFS0s$\_$HE (23 galaxies) and SFS0s$\_$LE (48 galaxies). Our analysis comprises a comparison of global properties (e.g., S$\acute{\rm e}$rsic index, B/T) between the two samples. Subsequently, we computed the $CI$ parameters ($CI_{\rm H_{\alpha}/cont}$ and $CI_{\rm SFR, H_{\alpha}}$), followed by the construction of radial profiles for derived quantities of galaxies (e.g., rsSFR, D$_{\rm n}$4000). The key outcomes of our investigation are outlined below.

\begin{itemize}
\item By comparing the global properties of the two samples, we suggest that different environments do not significantly affect them. These properties changes in the total sample are consistent with previous works. Furthermore, the difference in these properties from \citet{2022MNRAS.515..201C} should be attributed to the different distribution of the SFR. This population is special and rare, which may only represent a very special phase in the transition from gas-rich SGs to gas-poor S0s.

\item The dominant perturbation processes in different environments have different effects on the gas and stars in the galaxy, so we first consider the $CI_{\rm H_{\alpha}/cont}$ parameter that includes the stellar and gas of the galaxy. We find that HDEs show a smaller $CI_{\rm H_{\alpha}/cont}$, while LDEs show a larger $CI_{\rm H_{\alpha}/cont}$. This is because the inflow of fresh gas from outside causes the distribution of gas in SFS0s$\_$LE to be more extended compared to the stellar disk. Galaxies that are subject to disturbances similar to RPS will experience the shrinking effect of H\,{\sc i} disk \citep[][]{2023ApJ...956..148L}. To confirm this result, we used $CI_{\rm SFR, H_{\alpha}}$, a parameter that only considers the gas disk of the galaxy. We find that it is indeed the merging or inflow of external gas in LDEs that causes the $CI_{\rm SFR, H_{\alpha}}$ value to be larger, while the stripping of gas in HDEs makes it smaller. We note that these differences are statistically significant.

\item We constructed radial profiles of some quantities ($\Sigma_{\rm SFR}$, rsSFR, D$_{\rm n}$4000, and gas-phase metallicity). We find that, on average, SFS0s$\_$LE has a higher $\Sigma_{\rm SFR}$/rsSFR, which is attributed to the injection of fresh external gas. These newly entered gases may not immediately participate in SF, only igniting the SF of the galaxy when the gas loses its original angular momentum and cooling \citep[][]{2019ApJ...883L..36H}. This is consistent with the origin mechanism of a ``minor merger." However, the disturbance mechanisms of SFS0s$\_$HE either force the gas in the outer disk to be stripped or stop the supply of gas for the galaxy, so a low $\Sigma_{\rm SFR}$/rsSFR is understandable. This is consistent with the ``faded spiral" origin. In addition, the central peak $\Sigma_{\rm SFR}$ of both samples does indicate that their SF activities are concentrated in the galaxy's center.

\item We find that it may be due to the inflow of external fresh gas into SFS0s$\_$LE that they have a lower and relatively flat age gradient of SPs (i.e., D$_{\rm n}4000$), while gas stripping or cessation of gas supply leads to a gradually increasing age gradient in SFS0s$\_$HE. This difference corresponds well to their different origin mechanisms. Most of this fresh gas from the outside is metal-poor, but we do not find a significantly lower gas-phase metallicity in SFS0s$\_$LE. The inspection of the measurements from the central 3$^{\prime \prime}$ within galaxies provided by MPA-JHU and the M$_{\ast}$ both show the same results. In fact, the gas-phase metallicity is the result of competition between gas inflow, outflow, supernova feedback, and SF feedback. It is only when the inflow of metal-poor gas is strong and SF does not last long enough to enrich the interstellar medium (ISM) through supernova explosions that the galaxy may exhibit a metallicity deficiency.
\end{itemize}

It is important to note that the results presented here are based on our analysis of SFS0s from the SDSS-IV MaNGA Survey. The fraction of high- and low-mass galaxies within our sub-sample is fairly balanced. However, given the relatively small total sample size (71), future investigations ought to explore the potential impact of M$_{\ast}$ on the SFS0s population. Considering the size of our sample and the unique nature of SFS0s, additional atomic and molecular gas data may provide a better understanding of this work \citep[][]{2023RAA....23a5005C, 2023MNRAS.523.1140L}. Deep H\,{\sc i} data, for instance, prove valuable in unraveling the historical and ongoing effects of gravitational and hydrodynamic processes, especially in transitioning galaxies from LDEs to HDEs \citep{2023MNRAS.523.1140L}. Most importantly, atomic and molecular gas data can help us answer some of the questions above and potentially determine the position of blue ETGs (e.g., SFS0s) in the overall evolutionary picture (Chen et al in prep). Furthermore, exploring feedback processes in S0s, such as AGN feedback, as well as understanding the progenitors \citep[e.g., passive and red SGs,][]{2019ApJ...880..149P, 2021ApJ...916...38Z} of these galaxies, is still of interest. The existence of passive and red SGs serves as compelling evidence for the morphological transformation of SGs into S0s, particularly in galaxy groups or rich galaxy clusters \citep[e.g.,][]{2019ApJ...880..149P}.


\begin{acknowledgements}
We are grateful to the anonymous referee for her/his thoughtful review and very constructive suggestions, which greatly improved this paper. We thank Dr. Qiusheng Gu and Dr. Min Du for the helpful discussion on related observations and simulations. J.W. acknowledges the National Key R\&D Program of China (Grant No. 2023YFA1607904) and the National Natural Science Foundation of China (NSFC) grants 12033004, 12221003, 12333002 and the science research grant from CMS-CSST-2021-A06. T.C. acknowledges the China Postdoctoral Science Foundation (Grant No. 2023M742929). This research made use of Photutils, an Astropy package for detection and photometry of astronomical sources \citep[][]{2022zndo...7419741B}. 

Funding for the Sloan Digital Sky Survey IV has been provided by the Alfred P. Sloan Foundation, the U.S. Department of Energy Office of Science, and the Participating Institutions. SDSS-IV acknowledges support and resources from the Center for High-Performance Computing at the University of Utah. The SDSS website is \url{www.sdss.org}. SDSS-IV is managed by the Astrophysical Research Consortium for the Participating Institutions of the SDSS Collaboration including the Brazilian Participation Group, the Carnegie Institution for Science, Carnegie Mellon University, the Chilean Participation Group, the French Participation Group, Harvard-Smithsonian Center for Astrophysics, Instituto de Astrof\'{i}sica de Canarias, The Johns Hopkins University, Kavli Institute for the Physics and Mathematics of the Universe (IPMU) / University of Tokyo, Lawrence Berkeley National Laboratory, Leibniz Institut f\"{u}r Astrophysik Potsdam (AIP), Max-Planck-Institut f\"{u}r Astronomie (MPIA Heidelberg), Max-Planck-Institut f\"{u}r Astrophysik (MPA Garching), Max-Planck-Institut f\"{u}r Extraterrestrische Physik (MPE), National Astronomical Observatory of China, New Mexico State University, New York University, University of Notre Dame, Observat\'{o}rio Nacional / MCTI, The Ohio State University, Pennsylvania State University, Shanghai Astronomical Observatory, United Kingdom Participation Group, Universidad Nacional Aut\'{o}noma de M\'{e}xico, University of Arizona, University of Colorado Boulder, University of Oxford, University of Portsmouth, University of Utah, University of Virginia, University of Washington, University of Wisconsin, Vanderbilt University, and Yale University.

Facilities: {\emph Sloan}

Software: astropy \citet[][]{2013A&A...558A..33A, 2018AJ....156..123A, 2022ApJ...935..167A}; Tool for OPerations on Catalogues And Tables \citep[TOPCAT;][]{2005ASPC..347...29T}

\end{acknowledgements}




\bibliography{main}
\bibliographystyle{aa}

\onecolumn
\begin{appendix}
\section{Detailed information on our two samples}

\begin{table}[htbp]
\caption{Detailed Information of SFS0s$\_$HE}
\centering
\resizebox{\textwidth}{!}{
\begin{tabular}{cccccccccccccc}
    \hline\hline
    PLATE$\_$IFU & Redshift & R.A. (deg) & Decl. (deg) & log($M_{\ast}/M_{\odot}$) & log(SFR$_{\rm SED}$ ($M_{\odot}$yr$^{-1}$)) & S$\acute{\rm e}$rsic index & $eta\_k$ & $R_{\rm e} (\prime\prime)$ & B/T & $T\_Type$ & $P\_LTG$ & $P\_S0$ & EW$\_$H$_{\alpha}$ (\,\AA) \\
    (1) & (2) & (3) & (4) & (5) & (6) & (7) & (8) & (9) & (10) & (11) & (12) & (13) & (14) \\
    \hline
    10505-6104 & 0.0217 & 140.7277 & 33.1420 & 9.336 $\pm$ 0.032 & $-1.135$ $\pm$ 0.130 & 0.21 $\pm$ 0.03 & 0.93 & 3.28 & 0.46 & $-2.288$ $\pm$ 0.570 & 0.48 & 0.79 & $-25.86$ $\pm$ 6.42 \\
    11759-3703 & 0.0201 & 146.8423 & 0.7408 & 9.275 $\pm$ 0.058 & $-0.506$ $\pm$ 0.102 & 2.15 $\pm$ 0.58 & 2.38 & 3.93 & 0.27 & $-1.143$ $\pm$ 0.661 & 0.02 & 0.57 & $-31.06$ $\pm$ 2.37 \\
    11761-6104 & 0.0276 & 195.9200 & 53.7232 & 10.439 $\pm$ 0.021 & $-1.979$ $\pm$ 0.704 & 8.00 $\pm$ 0.48 & 1.73 & 6.39 & 0.43 & $-1.303$ $\pm$ 0.505 & 0.05 & 0.71 & $-10.48$ $\pm$ 1.18 \\
    11834-1902 & 0.0435 & 223.0809 & $-0.2692$ & 10.197 $\pm$ 0.027 & $-0.162$ $\pm$ 0.093 & 1.04 $\pm$ 0.04 & 0.20 & 2.11 & 0.49 & $-1.588$ $\pm$ 0.724 & 0.04 &  0.59 & $-21.30$ $\pm$ 1.47 \\
    12079-3704 & 0.0431 & 30.4308 & $-1.0698$ & 11.064 $\pm$ 0.016 & 0.208 $\pm$ 0.158 & 1.96 $\pm$ 0.08 & 1.70 & 5.05 & 0.54 & $-0.441$ $\pm$ 0.790 & 0.04 & 0.54 & $-7.16$ $\pm$ 0.99 \\
    7975-6104 & 0.0788 & 324.8916 & 10.4835 & 11.238 $\pm$ 0.034 & 0.650 $\pm$ 0.121 & 2.01 $\pm$ 0.06 & 0.83 & 4.04 & 0.67 & $-0.452$ $\pm$ 0.710 & 0.17 & 0.58 & $-11.81$ $\pm$ 2.57 \\
    8082-6101 & 0.0213 & 50.1452 & $-1.0959$ & 9.061 $\pm$ 0.070 & $-0.402$ $\pm$ 0.121 & 0.99 $\pm$ 0.06 & 2.59 & 5.55 & 0.29 & $-0.381$ $\pm$ 0.601 & 0.08 & 0.53 & $-8.02$ $\pm$ 1.95 \\
    8083-3702 & 0.0238 & 49.9732 & 0.3914 & 9.719 $\pm$ 0.042 & $-0.449$ $\pm$ 0.133 & 1.15 $\pm$ 0.05 & 1.10 & 3.01 & 0.64 & $-0.106$ $\pm$ 0.630 & 0.24 & 0.60 & $-22.02$ $\pm$ 2.39 \\
    8097-3704 & 0.0258 & 27.2633 & 12.8736 & 9.552 $\pm$ 0.041 & $-0.596$ $\pm$ 0.126 & 1.51 $\pm$ 0.44 & 0.61 & 2.53 & 0.68 & $-0.953$ $\pm$ 0.610 & 0.09 & 0.71 & $-31.33$ $\pm$ 4.12 \\
    8149-3703 & 0.0267 & 119.3568 & 27.4403 & 9.765 $\pm$ 0.036 & $-1.358$ $\pm$ 0.427 & 4.14 $\pm$ 0.24 & 0.94 & 3.04 & 0.62 & $-0.258$ $\pm$ 0.600 & 0.02 & 0.56 & $-7.59$ $\pm$ 0.99 \\
    8249-3703 & 0.0264 & 139.7204 & 45.7278 & 9.873 $\pm$ 0.018 & 0.153 $\pm$ 0.023 & 1.00 $\pm$ 0.02 & 1.07 & 4.56 & 0.20 & $-2.339$ $\pm$ 0.459 & 0.02 & 0.57 & $-69.59$ $\pm$ 9.29 \\
    8262-3702 & 0.0242 & 183.6598 & 43.5362 & 10.264 $\pm$ 0.038 & 0.205 $\pm$ 0.061 & 1.06 $\pm$ 0.02 & 1.43 & 4.84 & 0.28 & $-0.122$ $\pm$ 0.400 & 0.09 & 0.56 & $-17.31$ $\pm$ 3.53 \\
    8336-3701 & 0.0179 & 210.3769 & 38.5159 & 9.157 $\pm$ 0.037 & $-1.018$ $\pm$ 0.075 & 6.29 $\pm$ 0.82 & 1.58 & 4.64 & 0.59 & $-1.058$ $\pm$ 0.564 & 0.45 & 0.80 & $-18.15$ $\pm$ 2.54 \\
    8588-1901 & 0.0364 & 249.7171 & 40.1993 & 9.924 $\pm$ 0.060 & $-0.304$ $\pm$ 0.186 & 0.19 $\pm$ 0.02 & 0.31 & 1.77 & 0.24 & $-1.150$ $\pm$ 0.628 & 0.20 & 0.90 & $-31.81$ $\pm$ 3.32 \\
    8622-3704 & 0.0402 & 351.2176 & 14.1389 & 10.075 $\pm$ 0.064 & $-0.005$ $\pm$ 0.016 & 5.58 $\pm$ 0.76 & 0.76 & 2.83 & 0.68 & $-0.228$ $\pm$ 0.499 & 0.04 & 0.53 & $-29.97$ $\pm$ 3.13 \\
    8715-6101 & 0.0543 & 119.1058 & 51.3446 & 10.789 $\pm$ 0.037 & 0.131 $\pm$ 0.080 & 2.21 $\pm$ 0.13 & 1.23 & 3.19 & 0.37 & $-2.069$ $\pm$ 0.630 & 0.03 & 0.61 & $-6.93$ $\pm$ 0.36 \\
    8985-3702 & 0.0238 & 204.0827 & 32.8751 & 9.655 $\pm$ 0.036 & $-0.445$ $\pm$ 0.076 & 2.02 $\pm$ 0.08 & 0.57 & 4.73 & 0.29 & $-0.921$ $\pm$ 0.760 & 0.09 & 0.54 & $-9.01$ $\pm$ 1.29 \\
    9029-6103 & 0.0304 & 246.7944 & 42.6789 & 10.099 $\pm$ 0.030 & $-0.538$ $\pm$ 0.158 & 6.66 $\pm$ 0.63 & 1.46 & 5.61 & 0.10 & $-3.359$ $\pm$ 0.563 & 0.43 & 0.73 & $-7.21$ $\pm$ 0.36 \\
    9037-3702 & 0.0188 & 234.8425 & 43.8654 & 9.741 $\pm$ 0.036 & $-0.44$ $\pm$ 0.125 & 5.73 $\pm$ 0.39 & 0.57 & 4.46 & 0.76 & $-2.220$ $\pm$ 0.521  & 0.02 & 0.58 &  $-18.75$ $\pm$ 2.45 \\
    9491-3701 & 0.0415 & 119.1066 & 17.9741 & 9.807 $\pm$ 0.038 & $-0.128$ $\pm$ 0.083 &  1.70 $\pm$ 0.09 & 1.74 & 2.61 & 0.52 & $-2.421$ $\pm$ 0.699 & 0.08 & 0.92 & $-17.92$ $\pm$ 2.07 \\
    9499-1901 & 0.0454 & 119.3980 & 25.8073 & 10.361 $\pm$ 0.030 & $-0.319$ $\pm$ 0.220 & 1.80 $\pm$ 0.09 & 1.36 & 1.42 & 0.60 & $-2.219$ $\pm$ 0.625 & 0.04 & 0.51 & $-9.16$ $\pm$ 0.32 \\
    9873-12702 & 0.0252 & 194.5855 & 27.4294 & 9.304 $\pm$ 0.041 & $-0.769$ $\pm$ 0.080 & 0.84 $\pm$ 0.03 & 2.37 & 3.54 & 0.30 & $-2.230$ $\pm$ 0.522 & 0.12 & 0.92 & $-14.14$ $\pm$ 1.39 \\
    9875-3704 & 0.0196 & 195.1152 & 27.6252 & 9.371 $\pm$ 0.047 & $-1.936$ $\pm$ 0.958 & 1.98 $\pm$ 0.06 & 2.22 & 3.34 & 0.75 & $-2.340$ $\pm$ 0.450 & 0.01 & 0.51 & $-71.61$ $\pm$ 5.56 \\
    \hline
\end{tabular}
}
\begin{tablenotes}
    \footnotesize
    \item[] Notes: The columns show (1) the MaNGA$\_$ID PLATE$\_$IFU of our targets; (2) Redshift; (3) R.A. (deg); (4) Decl. (deg); (5), (6) the stellar mass and SFR from \citet{2016ApJS..227....2S, 2018ApJ...859...11S}; (7) S$\acute{\rm e}$rsic index; (8) the projected number density parameter of the galaxy from GEMA$\_$VAC; (9) R$_{\rm e}$; (10) bulge-to-total light ratio; (11) $T\_Type$; (12) probability of LTG; (13) probability of S0; (14) the H$_{\alpha}$ equivalent width within the center of 2.5$^{\prime\prime}$. Pipe3D catalog defaults equivalent width to negative for emission lines \citep[][]{2020ARA&A..58...99S}.
\end{tablenotes}
\label{SFS0s_HE}
\end{table}

\begin{table}[htbp]
\caption{Detailed Information of SFS0s$\_$HE}
\centering
\resizebox{\textwidth}{!}{
\begin{tabular}{cccccccccccccc}
    \hline\hline
    PLATE$\_$IFU & Redshift & R.A. (deg) & Decl. (deg) & log($M_{\ast}/M_{\odot}$) & log(SFR$_{\rm SED}$ ($M_{\odot}$yr$^{-1}$)) & S$\acute{\rm e}$rsic index & $eta\_k$ & $R_{\rm e} (\prime\prime)$ & B/T & $T\_Type$ & $P\_LTG$ & $P\_S0$ & EW$\_$H$_{\alpha}$ (\,\AA) \\
        (1) & (2) & (3) & (4) & (5) & (6) & (7) & (8) & (9) & (10) & (11) & (12) & (13) & (14) \\
        \hline
        10218-1901 & 0.0222 & 118.0347 & 17.9303 & 9.423 $\pm$ 0.037 & $-0.163 \pm 0.029 $ & 0.62 $\pm$ 0.01 & $-999$ & 1.90 & 0.55 & $-0.635 \pm 0.521$ & 0.04 & 0.52 & $-85.24 \pm 16.16$\\
        10497-1901 & 0.0406 & 120.6247 & 17.2847 & 9.864 $\pm$ 0.025 & $-0.486 \pm 0.114 $ & 1.63 $\pm$ 0.11 & $-999$ & 1.71 & 0.74 & $-1.198 \pm 0.370$ & 0.29 & 0.86 & $-13.68 \pm 0.73$\\
        10841-3704 & 0.0227 & 142.3593 & 2.0707 & 9.580 $\pm$ 0.036 & $-0.833 \pm 0.189 $ & 1.71 $\pm $ 0.38 & $-999$ & 2.94 & 0.29 & $-0.450 \pm 0.340$ & 0.13 & 0.59 & $-6.14 \pm 0.55$\\
        11010-6101 & 0.0233 & 198.2229 & 23.6346 & 9.536 $\pm$ 0.047 & $-0.495 \pm 0.078 $ & 0.20 $\pm$ 0.01 & $-999$ & 5.62 & 0.26 & $-4.258 \pm 0.618$ & 0.10 & 0.97 & $-13.99 \pm 1.72$\\
        11753-3701 & 0.0933 & 146.6613 & 2.6589 & 11.452 $\pm$ 0.031 & $0.943 \pm 0.064 $ & 6.32 $\pm$ 1.00 & $-999$ & 2.77 & 0.58 & $-1.580 \pm 0.474$ & 0.02 & 0.60 & $-14.49 \pm 0.76$\\
        11759-3701 & 0.0457 & 145.5940 & 0.3094 & 10.505 $\pm$ 0.053 & $0.156 \pm 0.202 $ & 2.89 $\pm$ 0.32 & $-999$ & 2.68 & 0.44 & $-0.740 \pm 0.592$ & 0.15 & 0.75 & $-27.53 \pm 7.32$\\
        11823-1901 & 0.0368 & 248.2540 & 27.5934 & 9.41 $\pm$ 0.055 & $-0.275 \pm 0.059 $ & 1.10 $\pm$ 0.03 & $-999$ & 1.88 & 0.56 & $-1.468 \pm 0.584$ & 0.39 & 0.91 & $-51.74 \pm 10.46$\\
        11826-3702 & 0.0363 & 189.4614 & 38.8770 & 9.522 $\pm$ 0.043 & $-0.511 \pm 0.070 $ & 3.61 $\pm$ 0.39 & $-999$ & 2.37 & 0.70 & $-1.761 \pm 0.563$ & 0.37 & 0.79 & $-36.93 \pm 7.54$\\
        11838-1902 & 0.0346 & 156.7275 & $-0.5415$ & 11.321 $\pm$ 0.053 & $0.697 \pm 0.114 $ & 0.67 $\pm$ 0.02 & $-999$ & 3.39 & 0.33 & $-1.011 \pm 0.714$ & 0.01 & 0.46 & $-16.94 \pm 0.82$\\
        11948-1901 & 0.0364 & 250.2385 & 32.2889 & 9.605 $\pm$ 0.058 & $0.207 \pm 0.087 $ & 2.72 $\pm$ 0.18 & $-999$ & 1.69 & 0.52 & $-0.117 \pm 0.655$ & 0.02 & 0.61 & $-81.29 \pm 5.05$\\
        11952-12703 & 0.0338 & 254.4507 & 27.4168 & 10.413 $\pm$ 0.03 & $0.022 \pm 0.001 $ & 0.89 $\pm$ 0.02 & $-999$ & 10.64 & 0.31 & $-1.635 \pm 0.783$ & 0.12 & 0.67 & $-10.40 \pm 1.20$\\
        11975-1902 & 0.0358 & 252.8002 & 25.6509 & 9.512 $\pm$ 0.037 & $-0.385 \pm 0.035 $ & 2.23 $\pm$ 0.39 & $-999$ & 2.20 & 0.45 & $-1.879 \pm 0.569$ & 0.45 & 0.87 & $-39.95 \pm 3.85$\\
        12085-3701 & 0.0358 & 346.4024 & 13.6225 & 9.960 $\pm$ 0.051 & $-0.258 \pm 0.012 $ & 5.57 $\pm$ 2.04 & $-999$ & 2.13 & 0.66 & $-0.588 \pm 0.581$ & 0.05 & 0.67 & $-19.03 \pm 1.73$\\
        12089-1902 & 0.0171 & 352.3083 & 15.4361 & 9.328 $\pm$ 0.026 & $-0.718 \pm 0.018 $ & 0.94 $\pm$ 0.05 & $-999$ & 3.56 & 0.19 & $-3.627 \pm 0.600$ & 0.15 & 0.95 & $-19.80 \pm 1.71$\\
        12483-1901 & 0.0223 & 185.9798 & 33.9287 & 9.259 $\pm$ 0.044 & $-0.685 \pm 0.093 $ & 2.38 $\pm$ 0.21 & $-999$ & 1.90 & 0.54 & $-0.934 \pm 0.483$ & 0.18 & 0.58 & $-22.94 \pm 1.98$\\
        8138-3704 & 0.0263 & 118.0205 & 44.1615 & 9.995 $\pm$ 0.027 & $-0.310 \pm 0.091 $ & 0.88 $\pm$ 0.03 & $-999$ & 2.21 & 0.46 & $-1.747 \pm 0.558$ & 0.02 & 0.43 & $-18.07 \pm 0.99$\\
        8140-1902 & 0.0407 & 117.7660 & 41.9735 & 9.926 $\pm$ 0.043 & $0.023 \pm 0.002 $ & 0.55 $\pm$ 0.05 & $-999$ & 1.79 & 0.38 & $-0.657 \pm 0.542$ & 0.36 & 0.78 & $-52.20 \pm 2.55$\\
        8155-1902 & 0.0234 & 53.7267 & $-0.7889$ & 9.479 $\pm$ 0.063 & $-0.431 \pm 0.062 $ & 0.90 $\pm$ 0.06 & $-999$ & 2.79 & 0.41 & $-1.211 \pm 0.512$ & 0.08 & 0.52 & $-40.74 \pm 3.28$\\
        8158-1901 & 0.0384 & 60.8593 & $-5.4918$ & 9.450 $\pm$ 0.061 & $-0.148 \pm 0.063 $ & 1.00 $\pm$ 0.08 & $-999$ & 1.93 & 0.57 & $-1.810 \pm 0.556$ & 0.30 & 0.84 & $-46.96 \pm 3.73$\\
        8241-1902 & 0.0383 & 125.2728 & 17.8198 & 9.653 $\pm$ 0.064 & $-0.181 \pm 0.058 $ & 0.93 $\pm$ 0.03 & $-999$ & 2.23 & 0.51 & $-1.049 \pm 0.646$ & 0.12 & 0.78 & $-50.47 \pm 5.33$\\
        8243-3701 & 0.0431 & 128.1644 & 53.2332 & 10.366 $\pm$ 0.014 & $-1.993 \pm 0.561 $ & 7.00 $\pm$ 0.43 & $-999$ & 2.73 & 0.73 & $-1.516 \pm 0.597$ & 0.02 & 0.45 & $-89.66 \pm 15.12$\\
        8259-1902 & 0.0242 & 177.7961 & 44.0088 & 9.854 $\pm$ 0.031 & $-0.407 \pm 0.100 $ & 0.88 $\pm$ 0.01 & $-999$ & 2.86 & 0.33 & $-0.176 \pm 0.434$ & 0.16 & 0.88 & $-17.37 \pm 2.17$\\
        8261-1901 & 0.0231 & 182.8574 & 44.4360 & 9.589 $\pm$ 0.043 & $-0.332 \pm 0.073 $ & 0.36 $\pm$ 0.01 & $-999$ & 2.36 & 0.59 & $-0.484 \pm 0.577$ & 0.02 & 0.59 & $-40.90 \pm 2.30$\\
        8313-6103 & 0.0238 & 239.4366 & 41.7095 & 9.906 $\pm$ 0.041 & $0.052 \pm 0.058 $ & 1.25 $\pm$ 0.02 & $-999$ & 3.37 & 0.33 & $-0.938 \pm 0.572$ & 0.03 & 0.58 & $-24.91 \pm 6.06$\\
        8314-3702 & 0.0302 & 240.8477 & 39.9857 & 10.267 $\pm$ 0.029 & $0.110 \pm 0.078 $ & 6.29 $\pm$ 0.83 & $-999$ & 4.08 & 0.59 & $-0.822 \pm 0.651$ & 0.03 & 0.49 & $-19.99 \pm 2.81$\\
        8315-9101 & 0.0626 & 235.7861 & 40.0486 & 11.105 $\pm$ 0.04 & $0.321 \pm 0.207 $ & 1.61 $\pm$ 0.68 & $-999$ & 4.19 & 0.32 & $-0.813 \pm 0.502$ & 0.31 & 0.78 & $-8.88 \pm 0.58$\\
        8323-3701 & 0.0332 & 195.3896 & 34.1167 & 9.616 $\pm$ 0.032 & $-0.696 \pm 0.148 $ & 0.98 $\pm$ 0.01 & $-999$ & 2.62 & 0.45 & $-0.842 \pm 0.503$ & 0.10 & 0.75 & $-12.74 \pm 1.78$\\
        8323-3703 & 0.0675 & 196.4398 & 34.6811 & 11.210 $\pm$ 0.027 & $0.557 \pm 0.099 $ & 1.78 $\pm$ 0.16 & $-999$ & 3.01 & 0.49 & $-0.227 \pm 0.649$ & 0.05 & 0.54 & $-7.13 \pm 0.88$\\
        8448-6101 & 0.0355 & 165.9620 & 23.0065 & 9.431 $\pm$ 0.045 & $-0.497 \pm 0.048 $ & 0.67 $\pm$ 0.02 & $-999$ & 3.35 & 0.46 & $-1.021 \pm 0.688$ & 0.02 & 0.54 & $-33.99 \pm 6.95$\\
        8455-1902 & 0.0262 & 155.7087 & 39.3689 & 9.626 $\pm$ 0.033 & $-0.429 \pm 0.093 $ & 0.68 $\pm$ 0.04 & $-999$ & 2.15 & 0.48 & $-1.397 \pm 0.437$ & 0.21 & 0.90 & $-41.25 \pm 2.81$\\
        8459-1902 & 0.0170 & 148.5025 & 43.0448 & 9.031 $\pm$ 0.042 & $-1.022 \pm 0.057 $ & 1.19 $\pm$ 0.02 & $-999$ & 3.11 & 0.38 & $-2.240 \pm 0.734$ & 0.38 & 0.82 & $-52.54 \pm 11.93$\\
        8462-3704 & 0.0398 & 143.3743 & 37.5026 & 9.986 $\pm$ 0.024 & $-0.380 \pm 0.068 $ & 0.28 $\pm$ 0.01 & $-999$ & 2.29 & 0.56 & $-1.127 \pm 0.471$ & 0.19 & 0.87 & $-15.41 \pm 2.60$\\
        8568-3703 & 0.0226 & 155.6931 & 37.6735 & 9.462 $\pm$ 0.074 & $0.034 \pm 0.006$ & 5.67 $\pm$ 0.62 & $-999$ & 4.40 & 0.45 & $-1.043 \pm 0.508$ & 0.10 & 0.55 & $-115.91 \pm 29.09$\\
        8652-1902 & 0.0460 & 331.0860 & $-0.5028$ & 10.289 $\pm$ 0.026 & $-0.027 \pm 0.094 $ & 1.53 $\pm$ 0.02 & $-999$ & 2.17 & 0.84 & $-0.448 \pm 0.477$ & 0.12 & 0.80 & $-15.70 \pm 2.39$\\
        8939-3701 & 0.0198 & 125.3617 & 24.4531 & 10.136 $\pm$ 0.034 & $-0.139 \pm 0.125 $ & 1.07 $\pm$ 0.01 & $-999$ & 5.26 & 0.38 & $-0.603 \pm 0.714$ & 0.03 & 0.53 & $-25.57 \pm 4.92$\\
        8940-6102 & 0.0276 & 120.8669 & 25.1027 & 10.540 $\pm$ 0.042 & $-0.156 \pm 0.012 $ & 1.49 $\pm$ 0.17 & $-999$ & 2.65 & 0.02 & $-0.639 \pm 0.657$ & 0.07 & 0.58 & $-45.34 \pm 5.12$\\
        8983-6102 & 0.0313 & 205.1829 & 25.9084 & 10.355 $\pm$ 0.029 & $-0.190 \pm 0.142 $ & 0.17 $\pm$ 0.03 & $-999$ & 7.13 & 0.37 & $-1.370 \pm 0.803$ & 0.44 & 0.78 & $-31.05 \pm 3.66$\\
        9035-12702 & 0.0383 & 236.5584 & 44.9646 & 9.964 $\pm$ 0.033 & $-0.208 \pm 0.098 $ & 8.00 $\pm$ 0.89 & $-999$ & 2.22 & 0.52 & $-1.568 \pm 0.700$ & 0.45 & 0.87 & $-24.93 \pm 2.10$\\
        9040-3703 & 0.0384 & 244.8737 & 28.5169 & 9.688 $\pm$ 0.061 & $-0.236 \pm 0.089 $ & 2.05 $\pm$ 0.23 & $-999$ & 2.57 & 0.09 & $-0.556 \pm 0.700$ & 0.06 & 0.59 & $-38.84 \pm 4.14$\\
        9092-1901 & 0.0344 & 240.6321 & 24.7622 & 9.488 $\pm$ 0.045 & $-0.444 \pm 0.081 $ & 2.71 $\pm$ 0.70 & $-999$ & 1.39 & 0.17 & $-0.029 \pm 0.489$ & 0.17 & 0.74 & $-23.85 \pm 2.41$\\
        9181-1902 & 0.0399 & 120.8730 & 38.4423 & 9.930 $\pm$ 0.024 & $-0.448 \pm 0.132 $ & 0.90 $\pm$ 0.06 & $-999$ & 1.85 & 0.83 & $-0.550 \pm 0.437$ & 0.15 & 0.80 & $-22.10 \pm 2.15$\\
        9183-6101 & 0.0225 & 122.0845 & 39.0236 & 10.267 $\pm$ 0.031 & $-0.028 \pm 0.119 $ & 0.50 $\pm$ 0.03 & $-999$ & 3.25 & 0.80 & $-0.240 \pm 0.614$ & 0.37 & 0.69 & $-28.62 \pm 1.31$\\
        9185-9101 & 0.0563 & 256.2123 & 34.8173 & 10.611 $\pm$ 0.063 & $1.488 \pm 0.053 $ & 1.93 $\pm$ 0.17 & $-999$ & 7.32 & 0.39 & $-0.367 \pm 0.652$ & 0.15 & 0.58 & $-96.38 \pm 9.18$\\
        9187-3704 & 0.0374 & 312.1987 & $-6.5032$ & 9.596 $\pm$ 0.043 & $-0.339 \pm 0.052 $ & 0.64 $\pm$ 0.03 & $-999$ & 2.34 & 0.32 & $-0.033 \pm 0.485$ & 0.10 & 0.52 & $-54.62 \pm 4.44$\\
        9488-1902 & 0.0182 & 127.0117 & 20.9967 & 9.072 $\pm$ 0.023 & $-0.741 \pm 0.012 $ & 1.55 $\pm$ 0.03 & $-999$ & 2.95 & 0.32 & $-1.123 \pm 0.620$ & 0.14 & 0.52 & $-36.61 \pm 3.96$\\
        9863-3703 & 0.0274 & 194.5132 & 26.9160 & 10.515 $\pm$ 0.02 & $-1.603 \pm 0.600 $ & 1.11 $\pm$ 0.06 & $-999$ & 4.05 & 0.34 & $-0.267 \pm 0.568$ & 0.05 & 0.76 & $-8.48 \pm 2.70$\\
        9865-1901 & 0.0134 & 222.4368 & 50.7109 & 9.592 $\pm$ 0.027 & $-0.420 \pm 0.148 $ & 1.10 $\pm$ 0.03 & $-999$ & 3.57 & 0.13 & $-0.918 \pm 0.703$ & 0.09 & 0.86 & $-24.78 \pm 3.74$\\
        9872-3701 & 0.0197 & 233.2320 & 42.4383 & 10.083 $\pm$ 0.043 & $0.012 \pm 0.057  $ & 0.63 $\pm$ 0.01 & $-999$ & 3.84 & 0.33 & $-0.418 \pm 0.644$ & 0.15 & 0.82 & $-32.23 \pm 5.75$\\
    \hline
\end{tabular}
}
\begin{tablenotes}
    \footnotesize
    \item[] \item[] Notes: All columns are the same as in Table \ref{SFS0s_HE}.
\end{tablenotes}
\label{SFS0s_LE}
\end{table}

\end{appendix}

\end{CJK}
\end{document}